\documentclass[fleqn,usenatbib]{mnras}

\usepackage{newtxtext,newtxmath}

\usepackage[T1]{fontenc}
\usepackage{ae,aecompl}

\usepackage{graphicx}	
\usepackage{amsmath}	
\usepackage{amssymb}	

\def\eg{{\it e.g.}}

\title[No pulsar left behind. I]{No Pulsar Left Behind. I. Timing, Pulse-sequence Polarimetry, and Emission Morphology for 12 pulsars.} 

\author[Brinkman et al.]{
Casey Brinkman,$^{1}$
\thanks{E-mail: clbrinkm@uvm.edu}
Paulo C. C. Freire,$^{2}$ 
Joanna Rankin$^{1,3}$ and
Kevin Stovall$^{4}$
\\
$^{1}$Physics Department, University of Vermont, Burlington, Vermont 05401, USA \\
$^{2}$Max-Planck-Institut f\"ur Radioastronomie, auf dem H\"ugel 69, 53121 Bonn, Germany \\
$^{3}$Anton Pannekoek Institute for Astronomy, University of Amsterdam, Science Park 904, 1098 XH Amsterdam\\
$^{4}$National Radio Astronomy Observatory, P.O. Box 0, Socorro, NM 87801, USA
}

\date{Accepted XXX. Received YYY; in original form ZZZ}

\pubyear{2015}

\begin{document}
\label{firstpage}
\pagerange{\pageref{firstpage}--\pageref{lastpage}}
\maketitle

\begin{abstract}
In this paper we study a set of twelve pulsars that previously had not been
characterized. Our timing shows that eleven of them are
``normal'' isolated pulsars, with rotation periods between 0.22 and 2.65 s,
characteristic ages between 0.25 Myr and 0.63 Gyr, and estimated magnetic fields
ranging from 0.05 to 3.8$\,\times\, 10^{12}\,$G. The youngest pulsar in our sample,
PSR~J0627+0706, is located near the Monoceros supernova remnant (SNR G205.5+0.5), but
it is not the pulsar most likely to be associated with it.
We also confirmed the existence of a candidate from an early Arecibo
survey, PSR~J2053+1718, its subsequent timing and polarimetry are also presented here.
It is an isolated pulsar with a spin period of
119 ms, a relatively small magnetic field of $5.8 \, \times \, 10^{9} \, \rm G$ and
a characteristic age of 6.7 Gyr;
this suggests the pulsar was mildly recycled by accretion from a companion star
which became unbound when that companion became a supernova.
We report the results of single-pulse and average Arecibo polarimetry at 
both 327 and 1400 MHz aimed at understanding the basic emission properties and 
beaming geometry of these pulsars. Three of them (PSRs~J0943+2253, J1935+1159 and J2050+1259) 
have strong nulls and sporadic radio emission, several others exhibit interpulses
(PSRs J0627+0706 and J0927+2345) and one shows regular drifting subpulses (J1404+1159). 
\end{abstract}




\section{Introduction and motivation}

\begin{table*} 
\begin{tabular}{|l| |c| |l| |l| |l| |r| |l| |r|} 
\hline 
Previous name  & Reference                  & New Name   & Start     & Finish    & NTOA  & rms   & Reduced \\ 
            &                            &            &  (MJD)       & (MJD)             &       & (ms)    & $\chi^
2$ \\
\hline
J0435+27      & \cite{1996ApJ...470.1103R} & J0435+2749 & 52854 & 53785   & 96    & 0.12  &  14.38    \\ 
J0517+22      & \cite{2003PhDT.........2C} & J0517+2212 & 53418 & 57701   & 518    & 0.12  &  7.9    \\ 
J0627+07*      & \cite{2003PhDT.........2C} & J0627+0706 & 53418 & 53675  & 91    & 0.26  & 22.82    \\ 
J0927+23      & \cite{1996ApJ...470.1103R} & J0927+2345 & 53318 & 53910  & 26    & 0.47  & 0.77    \\   
J0943+22      & \cite{1993ApJ...416..182T} & J0943+2253 & 53318 & 57876  & 731    & 0.17  &  2.57    \\ 
J0947+27      & \cite{1996ApJ...470.1103R} & J0947+2740 & 53318 & 53910  & 35    & 0.28  &  1.88    \\  
J1246+22      & \cite{1993ApJ...416..182T} & J1246+2253 & 53294 & 57876  & 448    & 0.22  &  4.28    \\ 
J1404+12      & \cite{2003PhDT.........2C} & J1404+1159 & 53309 & 57380  & 335    & 0.62  &  12.78    \\ 
J1756+18      & \cite{2003ApJ...594..943N} & J1756+1822 & 52645 & 53307  & 89    & 1.23  &  26.09    \\ 
J1935+12      & \cite{2003PhDT.........2C} & J1935+1159 & 53306 & 57504  & 145    & 2.51  &  1.04    \\ 
J2050+13      & \cite{2003ApJ...594..943N} & J2050+1259 & 52636 & 53306  & 27    & 8.81  & 1419.67    \\  
J2052+17**     & \cite{1996ApJ...470.1103R} & J2053+1718 & 53295 & 56837 & 728 & 0.028 & 1.34 \\
\hline 
\end{tabular}
\caption{Pulsar names (old and new), references and the parameters of our timing observations. *See also Burgay et al. (2013) **Previous unconfirmed candidate.}
\label{tab1}
\end{table*} 

Since the discovery of the first radio pulsar in 1967 \citep{1968Natur.217..709H}, more than 2500 rotation-powered pulsars have been discovered \citep{2005AJ....129.1993M}. Of these, more than 400 have been ``left behind''---that is, they have no published phase-coherent timing solutions, so that we lack a rudimentary knowledge of their proper motions, spin-down parameters (including characteristic age, magnetic field, spin-down luminosity) and possible orbital elements. Similarly, no polarimetry and
fluctuation-spectral analyses have been done for many pulsars, and for many of these even basic quantities such as flux densities, rotation measures and spectral indices are lacking.
Because of this, the scientific potential for many of these objects is simply unknown and  unexploited.

In this and subsequent papers, we attempt a partial remedy to this situation by 
characterizing some of these pulsars: we present their
timing solutions (with derivations of characteristic ages, surface magnetic field
and rotational spin-down) and study some of their radio emission properties.
As for all previously well characterized pulsars, the measurements presented here and
in subsequent papers will aid
future studies of the pulsar population and contribute to the understanding of their
emission physics.

In this first paper, we focus on a group of a dozen pulsars discovered with the 
430-MHz line feed of the Arecibo 305-m radio telescope in Puerto Rico {\em before}
the Arecibo upgrade, i.e., pulsars that were found 
more than 20 years ago, but were then never followed up. 
Most pulsars in this group were discovered in drift-scan surveys: two, J0943+22 and J1246+22, 
were reported by \cite{1993ApJ...416..182T} and three others (J0435+27, J0927+23, J0947+27) 
were discovered in the completion of that survey by \cite{1996ApJ...470.1103R}. 
Five further pulsars (J0517+22, J0627+07, J1404+12, J1935+12 and J1938+22) were
discovered in the Arecibo-Caltech drift-scan survey \citep{2003PhDT.........2C},
but again no timing solutions 
were presented for any of them.  Two of these pulsars were later timed
by other authors: J0627+0706, which we timed from  2005 Feb. 17 to Nov. 1,
was detected by the Perseus Arm pulsar survey and subsequently timed
from 2006 Jan 1 to 2011 May 9 \citep{2013MNRAS.429..579B}.
Their timing results are similar to ours, but more precise given the larger
timing baseline.
J1938+22, which we did not follow up, was later timed by
\cite{2013MNRAS.434..347L}, so that it is now known as J1938+2213.

Two other pulsars (J1756+18 and J2050+13) were discovered in the Arecibo 430-MHz 
intermediate latitude (pointed) survey \citep{2003ApJ...594..943N}.  They were reported 
in the above paper describing that survey, but without timing solutions because although
they were originally detected on the 19$^{\rm th}$ and 13$^{\rm th}$ of July 1990
respectively, they were confirmed only in January 2003.

\begin{figure*}
\centering
\includegraphics[width=\textwidth]{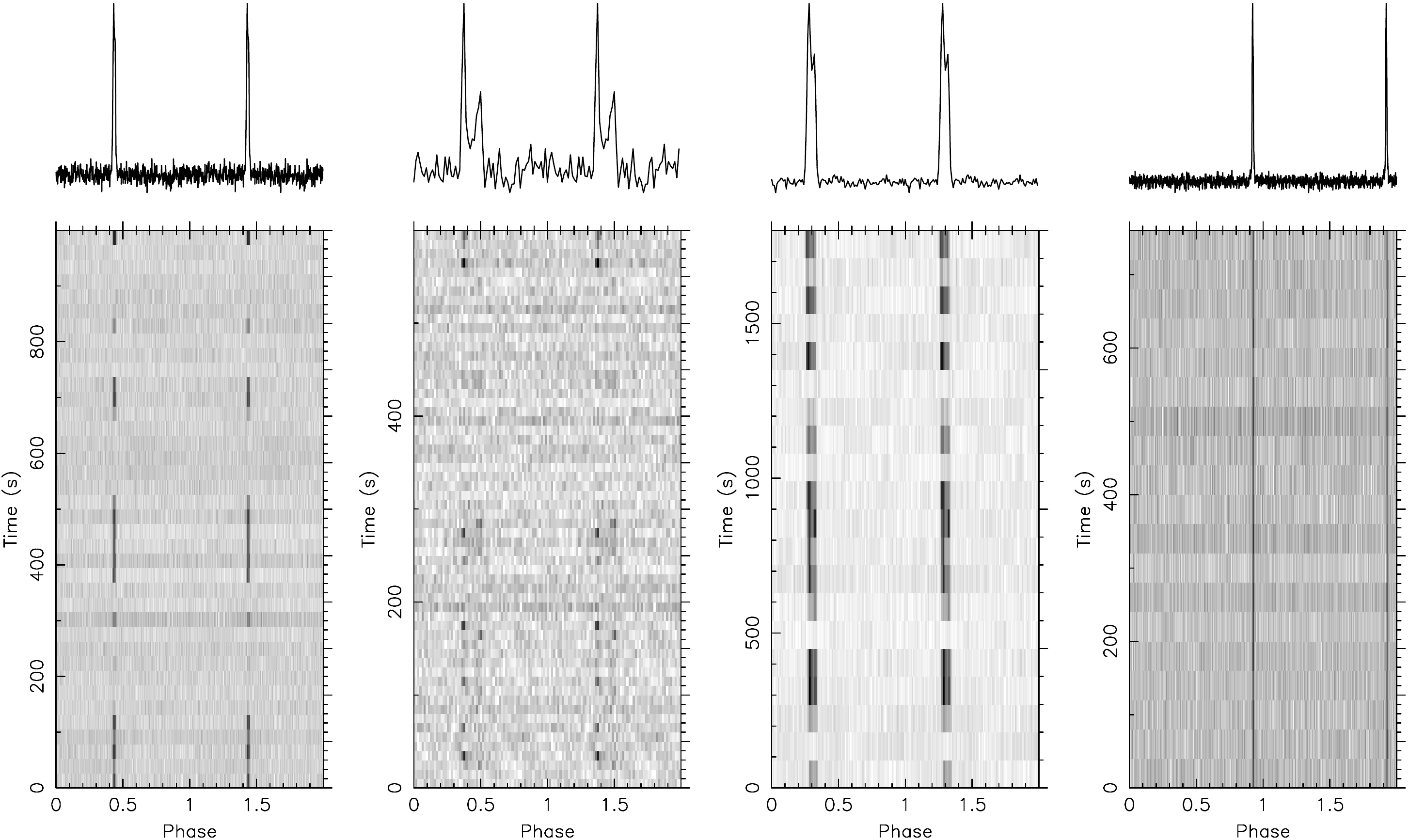}
\caption{Gray-scaled intensity plots (with darker shade implying a larger intensity)
showing total intensity of the pulsed signal at a frequency of 327 MHz as a function of
spin phase (2 spin cycles shown for clarity) and time. During many time segments
the pulsed signal of PSRs~J0943+2253, J1935+1159 and J2050+1259 (but not PSR~J2053+1718)
disappears, i.e., the first three pulsars null. The right plot shows the remarkably
narrow pulse profile of PSR~J2053+1718.} 
\label{fig2}
\end{figure*}

Finally, in \cite{1996ApJ...470.1103R} an additional pulsar candidate (J2052+17) was listed, but the
authors were unable to confirm it because of the start of the Arecibo upgrade. The
candidate had a spin period $P$ of 119.26 ms and a DM of $25 \pm 3 \, \rm cm^{-3}$\ 
pc. In 2004 October we confirmed the existence of this pulsar using the 327 MHz
Gregorian receiver of the 305-m Arecibo radio telescope and the Wideband Arecibo
Pulsar Processors (WAPPs, \citealt{2000ASPC..202..275D}) as back-ends. Both the topocentric spin
period (119.27 ms) and DM of $27 \, \rm cm^{-3}$\ pc were compatible with the
parameters in \cite{1996ApJ...470.1103R}. The pulsar has an exceptionally narrow profile,
which represents less than 1\% of a rotation cycle as presented in
Fig.~\ref{fig2} and in more detail in Fig.~\ref{figA12}.

In what follows, we present detailed studies for these objects, both of 
their timing and emission properties.  In \S\ref{sec:timing_observations} we discuss 
briefly the timing observations and their results,
\S\ref{sec:pol} describes the pulse-sequence, profile and polarization analyses, 
\S\ref{2053+1718} discusses the origin of PSR~J2053+1718
and \S\ref{disc} summarizes the various results.

\begin{table*} 
\begin{tabular}{l l l l l l} 
\hline
Pulsar Name &  Right Ascension  &  Declination    &   $\nu$   & $\dot{\nu}$              & DM             \\ 
   & hh mm ss.ss    & $^\circ \, \, \,   \arcmin \, \, \, \arcsec$  &  (Hz)              & ($10^{-16}$ Hz s$^{-1}$) & (pc cm$^{-3}$) \\
\hline 
J0435+2749  & 04 35 51.818(4)   & 27 49 01.7(4)  & 3.064857408039(13)   & $-$0.767(10)              & 53.19(2)  \\ 
J0517+2212  & 05 17 17.147(2) & 22 12 51.9(2)   & 4.497079963269(13)    & $-$2.3469(3)              & 18.705(14)  \\ 
J0627+0706  & 06 27 44.217(2)   & 07 06 12.7(11) & 2.1013960823(5)  & $-$1314.8(4)           & 138.29 \\ 
J0927+2345  & 09 27 45.26(5)      & 23 45 10.7(12) & 1.31252674625(15)   & $-$5.26(6)               & 17.24(12)  \\ 
J0943+2253  & 09 43 32.3975(10)     & 22 53 05.66(4) & 1.876261648496(3)  & $-$3.16238(7)          & 27.2508(15) \\ 
J0947+2740  & 09 47 21.287(18)    & 27 40 43.5(2)  & 1.17506907189(7)  & $-$5.94(3)             & 29.09(7)   \\ 
J1246+2253  & 12 46 49.363(5)   & 22 53 43.27(8)  & 2.110280928271(14) & $-$3.9586(4)               & 17.792(3) \\ 
J1404+1159  & 14 04 36.961(3)    & 11 59 15.36(10) & 0.377296025592(4) & $-$1.95656(16)             & 18.466(9) \\ 
J1756+1822  & 17 56 17.583(8)     & 18 22 55.3(2) & 1.34408432377(12)   & $-$9.27(3)             & 70.80  \\ 
J1935+1159  & 19 35 16.076(14)     & 11 59 09.2(4) & 0.51552817782(4)  & $-$2.5190(15)               & 188.76(6)  \\ 
J2050+1259  & 20 50 57.21(14)      & 12 59 09(3) & 0.8189874162(6)  & $-$3.38(15)               & 52.40  \\ 
J2053+1718  & 20 53 49.4809(7) & 17 18 44.662(13) & 8.384495643240(8) & $-$0.2014(7)      & 26.979 \\
\hline
\end{tabular}
\caption{Parameters from the timing solution for the reference epoch MJD = 53400. The digits in parentheses
indicate the 1-$\sigma$ uncertainty estimated by {\tt tempo} on the last digit
of the value.}
\label{tab2}
\end{table*} 

\begin{table*} 
\begin{tabular}{l r r l l l l l l l} 
\hline
Pulsar Name& \multicolumn{2}{|l|}{Galactic Coord.} & $P$    & $\dot{P}$     & $\log_{10}(\tau_c)$      & $\log_{10}(B_0)$          & $D_1$ & $D_2$ & $\log_{10} \dot{
E}$ \\
           &    \multicolumn{1}{|l|}{$\ell$}    &    \multicolumn{1}{|l|}{$b$}        & (s)       & ($10^{-15}\,$s s$^{-1}$ ) & ($\tau_c$ in yr) & ($B_0$ in G) & (kpc) & (kpc)  & ($\dot{E}$ in erg s$^{-1}$)\\
\hline
J0435+2749 & 171.8  & $-$13.1 & 0.3262794534509(14)    & 0.00816(11)    & 8.8        & 10.7        & 1.8 & 1.5   & 31.0 \\
J0517+2212 & 182.2  & $-$9.0 & 0.2223665151983(6)    & 0.0116045(15)   & 8.5        & 10.7        & 0.66 & 0.16    & 31.5 \\
J0627+0706 & 203.9  & $-$2.0 & 0.47587411455(11)     & 29.775(9)   & 5.4     & 12.6         & 4.7 & 2.3    & 34.0 \\
J0927+2345 & 205.3 & +44.2   & 0.76188923606(9)      & 0.305(4)    & 7.6       & 11.7         & 0.66 & 1.1   & 31.4 \\
J0943+2253 & 207.9  & +47.5   &  0.5329747057409(7)    &  0.089831(2)  & 8.0       & 11.3         & 1.2  & 3.5  & 31.4 \\
J0947+2740 & 201.1 & +49.4  & 0.85101380330(5)    & 0.430(2)    & 7.5       & 11.8         & 1.28 & *   & 31.4 \\
J1246+2253 & 288.8  & +85.6   & 0.473870557518(3)   & 0.088891(9)   & 7.9       & 11.3         & 1.5 & 2.5   & 31.5 \\
J1404+1159 & 355.1  & +67.1   & 2.65043873291(3)    & 1.37445(11)     & 7.5       & 12.3         & 1.4 & 2.2   & 30.4 \\
J1756+1822 &  43.8  & +20.2   & 0.74400093976(6)  & 0.5129(16)     & 7.4       & 11.8         & 4.2 & *   & 31.7 \\
J1935+1159 &  48.6  & $-$4.1 & 1.93975818009(15)     & 0.9478(6)       & 7.5        & 12.1        & 6.8 & 8.7   & 30.7 \\
J2050+1259 &  59.4  & $-$19.2 & 1.2210199817(9)  & 0.50(2)      & 7.6       & 11.9         & 3.1 & 5.9  & 31.0 \\
J2053+1718  & 63.6 &  $-$17.3   & 0.11926775831845(11) & 0.0002864(10) & 9.8     & 9.8   &   1.9 & 2.1 &  30.8 \\ 
\hline
\end{tabular}
\caption{Derived parameters. The digits in parentheses
indicate the 1-$\sigma$ uncertainty estimated by {\tt tempo} on the last digit
of the value. $D_1$ is calculated using the NE2001
model, $D_2$ is calculated using the YMW16 model. The asterisks indicate that the DM is larger than the model prediction for the total Galactic column density for the
pulsar's line of sight.}
\label{tab3}
\end{table*}

\section{Timing observations and results}   
\label{sec:timing_observations}

The aforementioned pulsars are relatively bright, and for that reason were used as 
test pulsars during the very early demonstration stages of the Arecibo 327-MHz drift scan survey 
(AO327, \citealt{2013ApJ...775...51D}).  These observations were carried out with the 327-MHz
Gregorian feed and one of the four WAPPs as backends.  
They all were acquired in search mode, with 256 spectral channels, a sampling time of 64$\mu$s and a 
bandwidth of 50 MHz.  They were later dedispersed and folded using the PRESTO 
routine ``prepfold'' \citep{2002AJ....124.1788R}, and then topocentric pulse times of arrival (TOAs) 
were derived from the resulting profiles using the FFT technique described by
\cite{1992RSPTA.341..117T} and implemented in the PRESTO routine get\_TOAs.py.

At a later phase (2015/2016), we have used some of these pulsars
(J0517+2212, J0943+2253, J1246+2253, J1404+1159 and J1935+1159) as test pulsars during follow-up sessions of AO327 discovered pulsars. These observations were made using the same 327-MHz Gregorian feed, but with the Puerto Rican Ultimate Pulsar Processing Instrument (PUPPI), which is a clone of the Greenbank Ultimate Pulsar Processing Instrument (GUPPI)\footnote{\url{https://safe.nrao.edu/wiki/bin/view/CICADA/GUPPiUsersGuide}}. Initial observations consisted of incoherent search mode observations with 69 MHz of bandwidth that was split into 2816 channels with a sample time of 81.92 $\mu s$. These data were dedispersed and folded using the {\tt fold\_psrfits} routine from the {\tt psrfits\_utils} software package\footnote{\url{http://github.com/scottransom/psrfits\_utils}}. Later observations were performed using PUPPI in coherent fold mode with the same 69 MHz of
bandwidth split into 44 frequency channels and were written to disk every 10 s. RFI was excised from both incoherent and coherent PUPPI files using a median-zapping algorithm included in the {\tt PSRCHIVE} software package \citep{2012AR&T....9..237V}\footnote{\url{http://psrchive.sourceforge.net/}} and topocentric TOAs were derived using PSRCHIVE's {\tt pat} tool.

\begin{figure*}
	\includegraphics[width=0.85\textwidth]{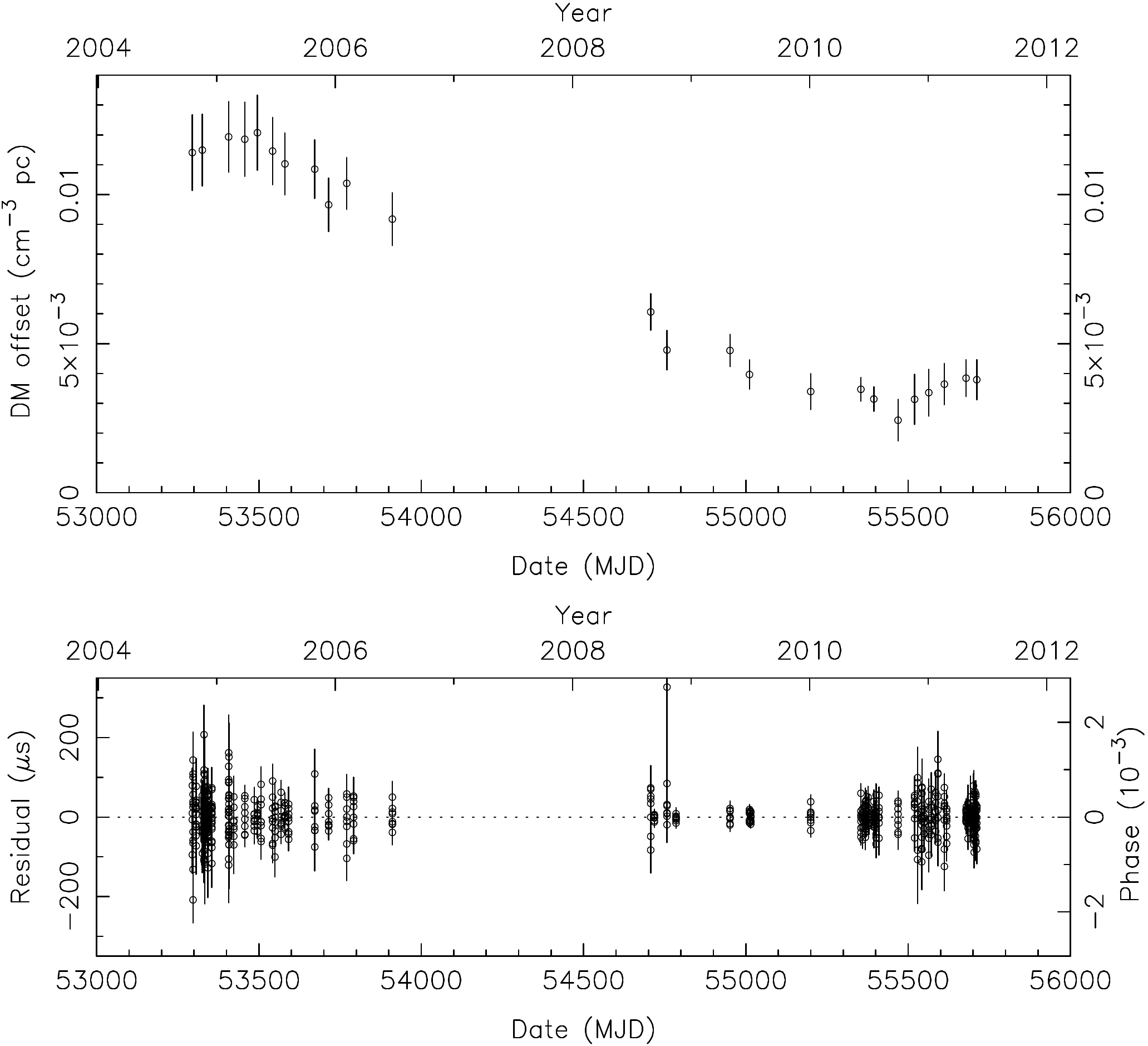}
    \caption{{\em Top plot}: Dispersion measure offset
(from 26.979 cm$^{-3}\,$pc) observed as a function of epoch for
PSR~J2053+1718. {\em Bottom plot}:
Residuals as a function of epoch for the same pulsar.
No trends are noticeable in the timing.}
    \label{fig:residuals}
\end{figure*}

The TOAs were then analyzed using {\tt tempo}\footnote{\url{http://tempo.sourceforge.net/}}, with the DE405 solar system
ephemeris\footnote{Standish, E.M.: 1998, "JPL Planetary and Lunar Ephemerides,
DE405/LE405", \url{ftp://ssd.jpl.nasa.gov/pub/eph/planets/ioms/de405.iom.pdf}};
from this we derive a timing solution, where we can assign the correct integer
rotation number to each pulse.

The characteristics of the timing observations are given in Table~1, together 
with the number of TOAs derived for each, the root mean square (rms) of the residuals 
(a residual is the measured TOA minus the prediction of the timing solution
for the time of the respective pulse). As an example, we depict graphically the residuals
of PSR~J2053+1718 in Fig. \ref{fig:residuals}) and the reduced $\chi^2$ of each fit.
For some pulsars this is much larger than 1, implying either the presence of effects
that have not been modelled in their timing solutions, such as timing noise or pulse
jitter, or that for some reason the TOA uncertainties were greatly under-estimated. 
The table also gives the new names for these 
pulsars, these names are used throughout the remainder of this paper.

The timing solutions themselves are presented in Table~2, these include
precise measurements of Right Ascension and Declination
(this is the origin of the new names in the preceding table)
and in some cases the proper motion along these two directions ($\mu_{\alpha}$ and $\mu_{\delta}$),
the rotation frequency ($\nu$), its derivative ($\dot{\nu}$), and dispersion
measure (DM). The reference epoch for all timing parameters is MJD = 53400.
For these measurements, we have multiplied the TOA uncertainties by the square root
of the reduced $\chi^2$ in Table 1, this yields a new reduced $\chi^2$ of 1.0.
This procedure results in conservative estimates of the uncertainties of the timing parameters.
These might still be in error due to the presence of correlated noise in the TOAs,
such as that produced by timing noise; this is particularly important for
PSRs J0627+0706, J1756+1822 and especially PSR J2050+1259.

Finally, the derived parameters are given in Table~3: Galactic coordinates ($l,b$) 
and distance ($D$), the latter is derived from the DM using two models of the
electron distribution in the Galaxy (the NE2001 model, \citealt{2002astro.ph..7156C}
and the YMW16 model, \citealt{2017ApJ...835...29Y}), the spin period ($P$), its
derivative ($\dot{P}$), the characteristic age ($\tau_c$), magnetic field ($B_0$), and 
spin-down energy ($\dot{E}$).  The expressions for these quantities were adopted from 
\cite{2004hpa..book.....L}:  $\tau_c = P / (2 \dot{P})$, $B = 3.2 \times 10^{19} \sqrt{P \dot{P}}$ 
and $\dot{E}\, =\, 4 \pi^{2} \, I \, \dot{P}/P^{3}$, where $I$ is the moment of inertia of the 
neutron star, which is generally assumed to be $10^{45}\, \rm g \, cm^{2}$.  All pulsars in 
the list are isolated, and most of them belong to the ``normal'' group, with fairly typical 
rotation periods (between 0.22 and 2.65 s), characteristic ages (between 0.25 Myr and 
0.63 Gyr) and B-fields (from 0.05 to 3.8$\,\times\, 10^{12}\,$G).
We now discuss the characteristic of the two extreme objects in our sample.

\subsection{PSR~J0627+0706}

With a characteristic age of only 250 kyr, PSR~J0627+0706 is
by far the most energetic object in this sample.
This object displays a prominent interpulse, 
suggesting an orthogonal rotator. It also has significant timing noise, which is
typical of young pulsars. This is the reason for the large reduced
$\chi^2$ of its solution. This agrees with the results
of \cite{2013MNRAS.429..579B}.

\cite{2003PhDT.........2C} remarked that this pulsar lies, in projection,
within $3^\circ$ of the centre of the old, large Monoceros supernova remnant (SNR G205.5+0.5);
which is located at an RA of $06^{\rm h} \, 39^{\rm min}$
and Declination of $\, +06^\circ\, 30\arcmin$.
For this reason they proposed that, if PSR~J0627+0706 has a true age similar to that
of SNR G205.5+0.5 (30 to 100 kyr, \citealt{1986MNRAS.220..501L}), it could be a
candidate in association with that SNR (they point out that the positional offset
is possible given the age of the SNR and the proper motions of other pulsars
observed in the Galaxy).

The characteristic age we and \cite{2013MNRAS.429..579B}
measure for PSR~J0627+0706 is larger by a factor of 2.5 to 8. However, this is not
very constraining: true pulsar ages can be significantly smaller than their characteristic ages
if they are born with a spin period similar to their current spin period.

Another way of verifying the association is through distance measurements.
The estimated distance to the Monoceros SNR, 1.6 kpc \citep{1986ApJ...301..813O},
is smaller than the estimated DM distances to this pulsar, 4.5 kpc (NE 2001) and 2.3 kpc
(YMW16).
Given the uncertainties in the DM models and their derived distances, this does not imply
a distance inconsistency, particularly for the YMW16 model.

We note that this area of the Galaxy has an abundance of 
relatively young pulsars that could potentially be associated with
SNR G205.5+0.5. In particular, the ``radio quiet'' gamma-ray pulsar PSR~J0633+0632, discovered
in data from the Fermi satellite \citep{2009Sci...325..840A}, was listed in
the latter paper as a ``plausible'' association with SNR G205.5+0.5
based on its location within $\sim 1.5^\circ$ of the SNR centre.
Later \cite{2011ApJS..194...17R} measured the characteristic age of PSR~J0633+0632 (59 kyr)
which is also more compatible with the estimated age of the SNR.
However, the latter pulsar has yet no detected radio emission, so its 
DM (and derived distance) is not yet known.

Despite that, we conclude that PSR~J0633+0632 is more likely
to be associated to SNR G205.5+0.5 than PSR~J0627+0706.

\subsection{PSR J2053+1718}

PSR~J2053+1718 has a much smaller spin period
derivative and a much larger characteristic age than the other pulsars in this sample.
A first simple fit for the parameters in Table~\ref{tab2}  plus proper motion (see below)
 yields a reduced $\chi^2$ of 1.80, and visible trends in
the residuals. Given the low frequency used in the timing and the relatively high
precision of the measurements, these are likely due to variations in DM caused by
the Earth's and the pulsar's movement through space. In order to
measure the DM variation with time, we divided the 50-MHz bandwidth in
4 sub-bands, making separate TOAs for each sub-band. We then used the DMX model
\citep{2013ApJ...762...94D} to measure the DM variations (displayed
in Fig.~\ref{fig:residuals}) and subtract them; once this is done the
reduced $\chi^2$ decreases to 1.34.

The precise timing of this pulsar allows a
measurement of its proper motion in right ascension and declination:
$\mu_{\alpha}\, = \, -1.0(23)\rm \, mas \, yr^{-1}$
and $\mu_{\delta}\, = \, +6.9(28)\rm \, mas \, yr^{-1}$.
These measurements are not highly significant, but they already represent
a significant constraint on the trasnverse velocity $v_{\rm T}$.
Given the DM distance estimates of 1.9 kpc (NE2001) and 2.1 kpc (YMW16),
the proper motion implies 
$v_{\rm T} \, = \, (63\, \pm \,  25)\, \rm km \, s^{-1}$
and $(69\, \pm \,  28)\, \rm km \, s^{-1}$ respectively; this is
typical among recycled pulsars (e.g., \citealt{2011ApJ...743..102G}).

This proper motion allows for a correction of the observed $\dot{P}$, where we
subtract the Shklovskii effect \citep{1970SvA....13..562S} and the Galactic
acceleration of this pulsar relative to that of the Solar System, projected along the
line of sight \citep{1991ApJ...366..501D} (this was calculated using the latest
model for the rotation of the Galaxy from 
\citealt{2014ApJ...783..130R}).  These terms mostly cancel each other, so that
the intrinsic spin-down, $\dot{P}_{\rm int} \, = \, 2.8 \times 10^{-19}\, \rm s \,
s^{-1}$, is very similar to the observed $\dot{P}$. This implies a low B-field of
$\sim \, 5.8\,  \times \, 10^9\, $G and a characteristic age of $\sim \,6.7\,
$Gyr. The origin of this pulsar is discussed in \S\ref{2053+1718}.

\section{Pulse-sequence and Profile Analyses and Quantitative Geometry}
\label{sec:pol}

Recently, we conducted single-pulse polarimetric observations on most of the above 
pulsars as well as a few others of related interest.  The Arecibo observations were 
carried out at both P band (327 MHz) and L band (1400 MHz) using total bandwidths 
of 50 MHz and typically 250 MHz, respectively.  Four Mock spectrometers were used 
to sample adjacent subbands after MJD 56300 (see \citealt{2016MNRAS.460.3063M}) and four 
Wideband Arecibo Pulsar Processors (WAPPs) earlier \citep{2013MNRAS.433..445R} to achieve 
milliperiod resolution.  The observations were then processed, calibrated and RFI excised (\eg, see \citealt{2012MNRAS.424.2477Y}) to provide pulse sequences that were used both to compute average polarization profiles and fluctuation spectra.  Rotation measures (RMs) were estimated for each of the pulsars by maximizing the linear polarization in the course of the polarimetric calibrations and analyses.  
     
A summary of these polarimetric observations are given in Table~4, as described above.  Nominal values of the rotation measure are also given in the table, and a complete description of the methods and errors will be given in a subsequent paper with many others.  Below 
we treat the various pulsars object by object referring to the polarized profiles and 
fluctuation spectra in the Appendix figures.  The analyses proceed from polarimetry 
to fluctuation spectra and finally to quantitative geometry following the procedures 
of \cite{1993ApJS...85..145R,1993ApJ...405..285R}.  The longitude-resolved fluctuation
(LRF) spectra of the pulse 
sequences (e.g., see \citealt{2001MNRAS.322..438D}) were computed in an effort to 
identify subpulse ``drift'' or stationary modulation associated with a rotating (conal) 
subbeam system.  

We have also attempted to classify the profiles where possible and conduct a quantitative geometrical analysis following the procedures of the core/double-cone model in Rankin (1993a,b; hereafter ET VI).  Outside half-power (3-dB) widths are measured for both conal components or pairs---and where possible estimated for 
cores.  Core widths can be used to estimate the angle between the rotation axis and magnetic axis $\alpha$,  polarisation position angle (PPA)-traverse central rates  $R=\sin\alpha/\sin\beta$ can be used to compute $\beta$ (the smallest 
angle between the magnetic axis and the line of sight),
and the conal widths can be used to compute 
conal beam radii using eqs. (1) through (6) of the above paper.  

The notes to Table~5 summarize our measurements, and the table values show the 
results of the geometrical model for the pulsar's emission beams.  The profile class 
is given in the first column, $\alpha$ and $\beta$ in next two per the $R$ value when 
possible.  The conal component profile widths $w$, conal beam radii $\rho$, and 
characteristic emission heights $h$ are tabulated in the rightmost three columns.  

\subsection{Analysis of Each Pulsar}

\begin{figure*}
   \centering
   \includegraphics[width=0.8\textwidth]{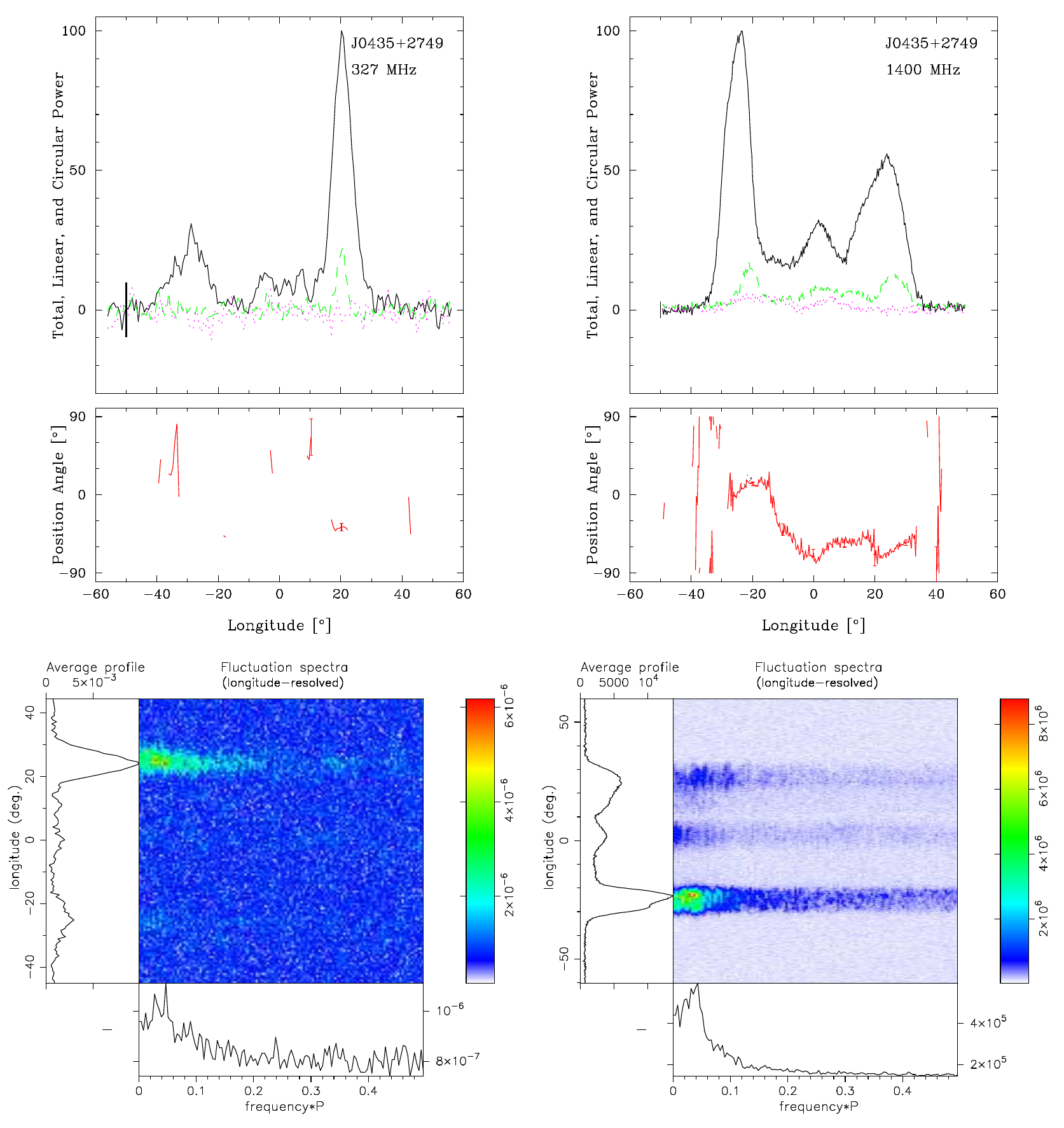}
   \caption{PSR J0435+2749 polarized profiles (upper displays) and fluctuation spectra 
(lower displays) at 327 MHz (left) and 1400 MHz (right).  The upper panels of the polarization displays give the total intensity in mJy (Stokes $I$; solid curve), the total linear ($L$ (=$\sqrt{Q^2+U^2}$); dashed green), and the circular polarization (Stokes $V$; dotted red){\bf ---with a righthand bar indicating 3-$\sigma$ in the off-pulse noise level}.  The PPA [=$(1/2)\tan^{-1} (U/Q)$] single values (dots, lower panels) in plots of the stronger pulsars correspond to those samples having errors smaller than 2\ $\sigma$ in $L$, and the average PPA is over plotted (solid red curve) with occasional 3-$\sigma$ errors. The longitude-resolved fluctuation spectra show the power levels {\bf (arbitrary units)} in the main panel {\bf (in cycles per rotation period)} according to the color bars (right). The average profiles are given in the left-hand panels and the aggregate fluctuating power in panels at the bottom of each display.} 
\label{figA1}
\end{figure*}

\noindent{\bf J0435+2749} has a clear triple profile at both frequencies as shown 
in Figure~\ref{figA1}, though the 1400 MHz profile is of better quality.  The leading and trailing components have very different spectral indices: the trailing component is much stronger at 327 MHz, but the leading is much stronger at 1400 MHz.  The fractional linear polarization is low in both profiles, and the PPA traverse is well defined only at the higher frequency. The power under the leading component may represent a different orthogonal polarization mode (OPM) than the others, suggesting little PPA rotation across the profile.  Both fluctuation spectra show broad peaks at about 0.05 cycles/period, primarily in the two outer components, which suggest a 20-period conal modulation. 

The observed conal spreading between the two frequencies suggests a core/outer cone beam geometry. The core width can only be estimated at the higher frequency at an upper limit of 9\degr\, implying that $\alpha$ is less than 1.4\degr\ per ET VIa, eq.(1).  

The PPA traverse shows a much more complex behavior than the Rotation Vector Model (RVM) describes; however, the roughly 90\degr\ rotation near the center of the pulse allows us to calculate the emission geometry using the RVM. The spherical geometric beam model in Table~5
seems compatible with a core-cone triple T classification.

\noindent{\bf J0517+2212} has a double component profile at 1400 MHz, whereas its  
327 MHz profile shows some structure in the second component. The polarization 
traverse shows little rotation at 327 MHz apart from the two 90\degr\ ``jumps'' and the 
behaviour seems similar at 1400 MHz but less well resolved--perhaps because of the 
diminished fractional linear polarization.  The fluctuation spectra shows a peak around 0.12 cycles/period, suggesting an 8 rotation-period modulation. 

The increased structure at 1400 MHz, the overall pulse narrowing, and the lack of PPA traverse suggest that $\beta\sim0$ and imply the outer conal beam geometry shown in Table~5.  
\begin{figure*}
   \centering
   \includegraphics[width=0.8\textwidth]{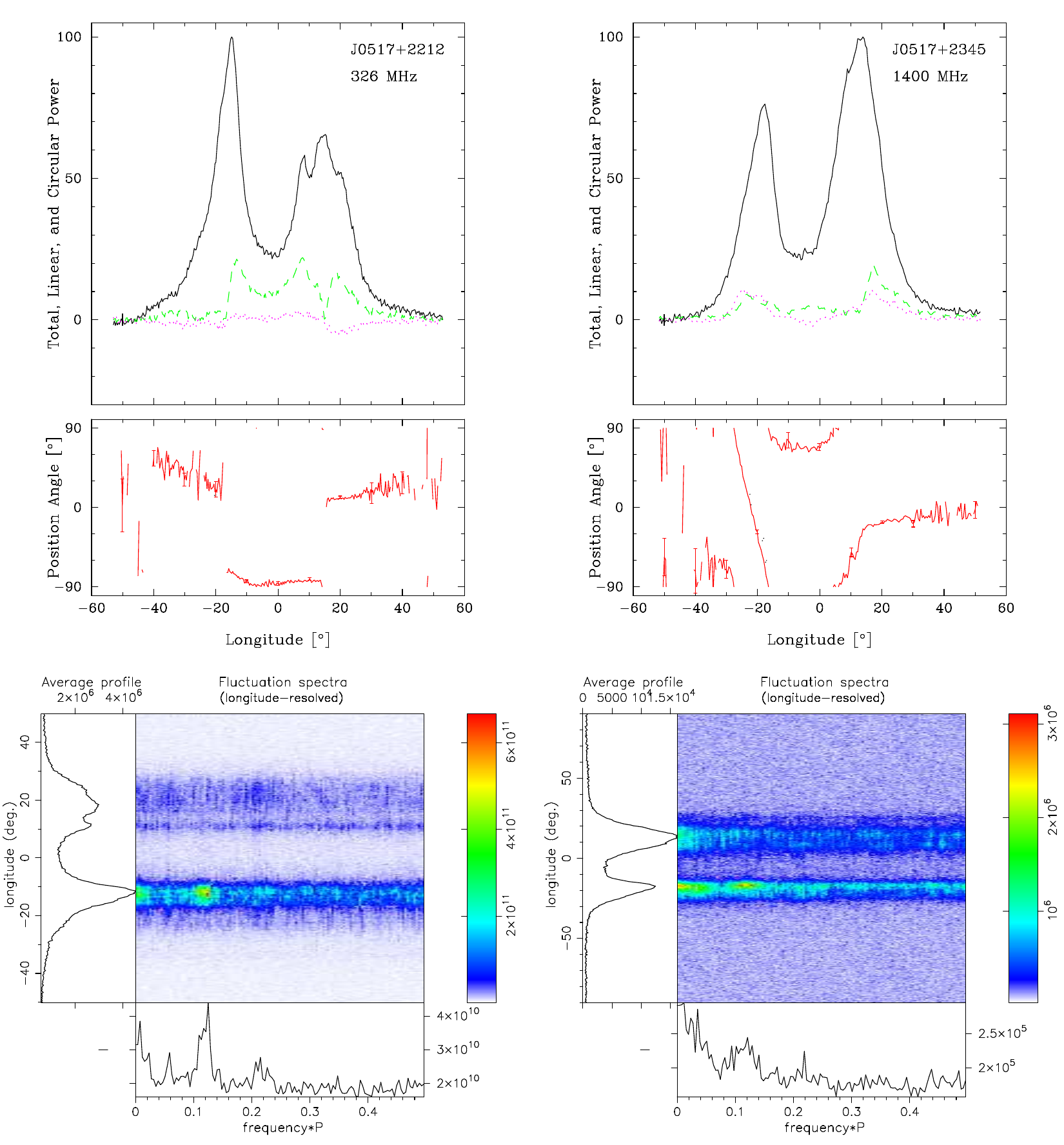}
\caption{J0517+2345 polarized profiles and fluctuation spectra as in Fig.~\ref{figA1}.}
\label{figA2}
\end{figure*}

\begin{figure*}

\includegraphics[width=\textwidth]{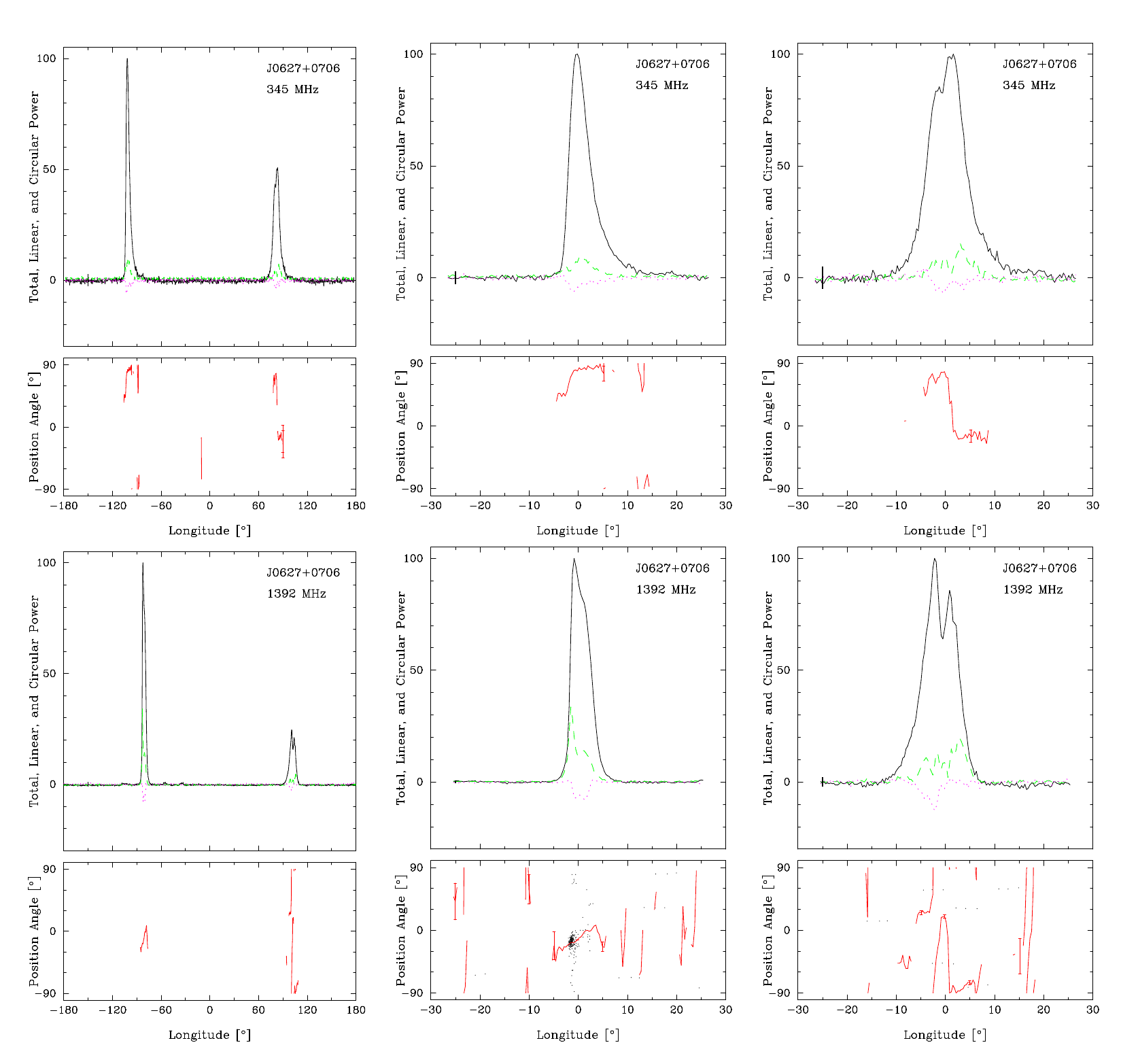}

\caption{J0627+0706 327- (upper row) and 1400-MHz (lower row) MHz polarized 
profiles as in Fig.~\ref{figA1}.  Full period displays on the left show the pulsar's 
main pulse and interpulse and the center and righthand plots show them separately.}
\label{figA3}
\end{figure*}

\noindent{\bf J0627+0706}  has a bright interpulse, separated from the main pulse by 177\degr\ ($\pm$ 0.27 \degr) as can be seen in Fig.~\ref{figA3}.  Because both features have structure, a more detailed interpretation is needed to assess how close to 180\degr\ they fall. However, given the narrowness of both features, it seems likely that they represent emission from the star's two poles, implying an orthogonal geometry where $\alpha$ is close to 90\degr.  PPA tracks give hints about the geometry only at 1400 MHz, and here little to go on apart from a probable 90\degr\ ``jump'' under the main pulse.  The main pulse might have three components and the interpulse two.  The fluctuation spectra are not displayed because they showed only flat ``white'' fluctuations. 

The pulse width at half maximum widens from 327 MHz to 1400 MHz, suggesting a core-single configuration where the conal emission is seen mainly at high frequencies. The polarization traverse is clearer at 1400 MHz, marked by prominent 90\degr\ modal ``jumps''.  It thus seems likely that $\beta$ is small for the main pulse and perhaps larger for the inter pulse, but that $\alpha$ is close to 90\degr, yielding the classifications and model values in Table~5. The emission heights given for both the main pulse and interpulse suggest that it is inner conal emission.

\begin{figure*}
\includegraphics[width=\textwidth]{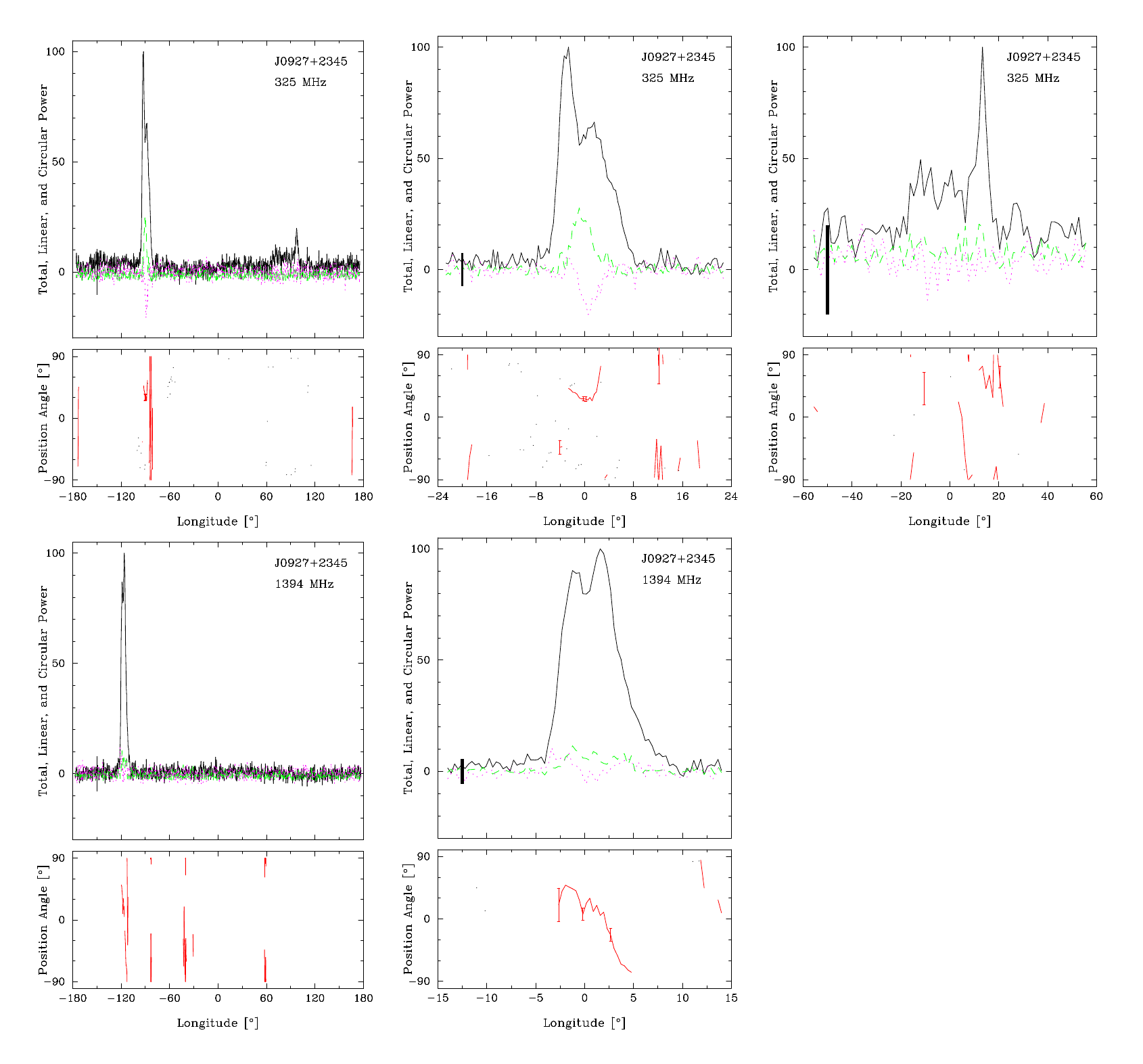}
\caption{J0927+2345 as in Fig.~\ref{figA3}. The pulsar's weak interpulse was 
not detected at the higher frequency separately.}
\label{figA4}
\end{figure*}

\noindent{\bf J0927+2345} shows an interesting feature at 327 MHz approximately 
180\degr\ away from the main pulse (Fig.~\ref{figA4}). This apparent interpulse is 
discernible only in the 327 MHz profile, and disappears in the 1400 MHz profile. Again, both profiles show so little linear polarization that a reliable PPA rate can be estimated for only a narrow longitude interval at 327 MHz. The main pulse appears to have three closely spaced features. The fluctuation spectra are not given for this pulsar because they showed no discernible features. 

The half width broadens from 327 MHz to 1400 MHz suggesting a core single evolution where conal ``outriders'' appear or become more prominent at high frequency.  The polarization traverse is well defined only for a short interval at 1400 MHz, which provides a useful $R$ value that leads to the geometric beam model in Table~5. If this profile were a triple, the central component would be $2.8\degr$ wide.

\noindent{\bf J0943+2253}  nulls as seen in Fig.~\ref{fig2} above, but the fluctuation spectra show no quasiperiodic behavior. The linear  polarization is slight and the PPA track shows what appears to be a 90\degr\ ``jump'' within the narrow interval where it is clearly defined. 

The average 327 MHz profile has two closely spaced components and perhaps an unresolved weak feature on its leading edge, suggesting a double or triple configuration. The 1400 MHz profile shows one main component, but there appears to be a bump on the leading edge indicating an additional component. The linear polarization is slight and difficult to interpret at 327 MHz, but is slightly more pronounced at 1400 MHz. We measure a central PPA rate of 7\degr/\degr\ at both frequencies in order to compute the geometric model parameters in Table~5, which suggest an inner-conal configuration. 
\begin{figure*}
\includegraphics[width=0.8\textwidth]{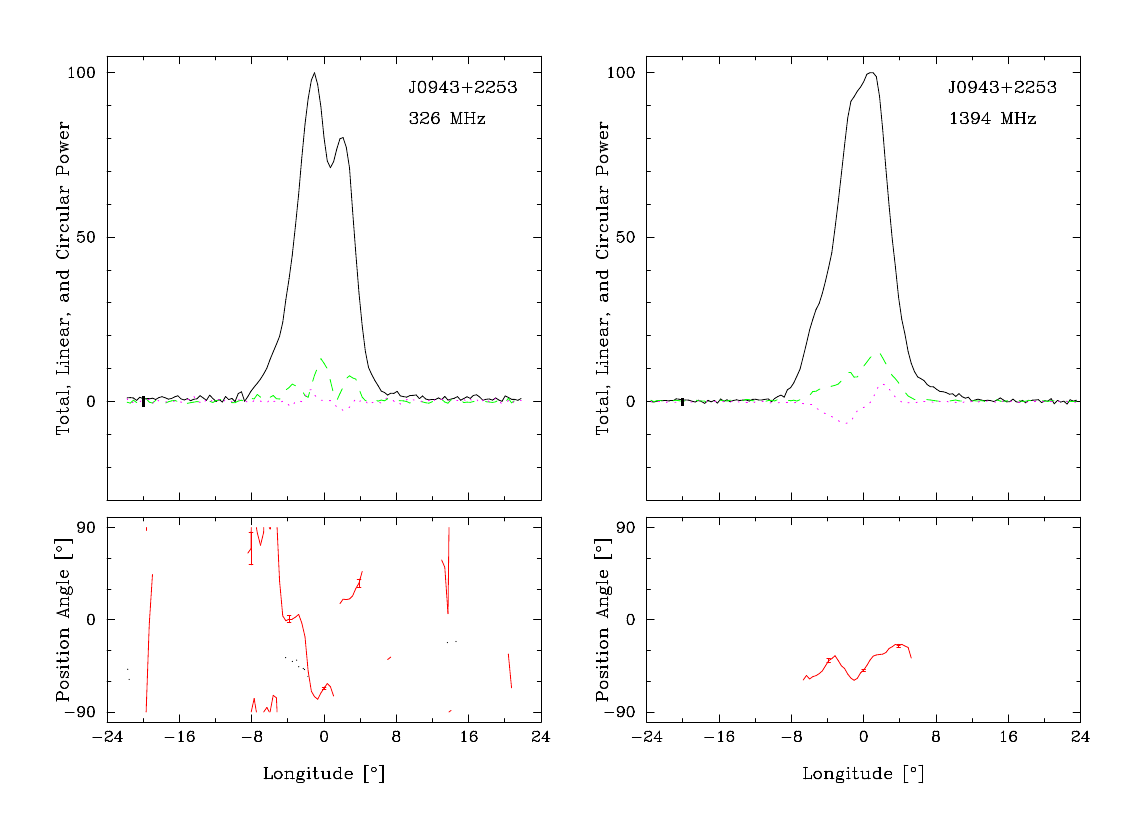}
\caption{J0943+2253 polarized profiles as in Fig.~\ref{figA1}.} 
\label{figA5}
\end{figure*}

\noindent{\bf J0947+2740} shows three components at both frequencies, though the 
leading region may be more complex at the lower frequency, probably representing 
a core and closely spaced inner cone. At 1400 MHz the conal components are weaker, and 
the PPA traverse is more complex. RVM behaviour and the slope of its polarization traverse could not be determined.  This pulsar is also known to exhibit sporadic emission between intervals of weakness or nulls as in Fig.~\ref{fig2}.  The fluctuation spectra (not shown) seem to hint at fluctuation power at periods longer than about 3 or 4 pulses.  

The fact that the profile width increases with wavelength indicates an outer-conal configuration. Its PPA traverse seems interpretable per the RVM model at 327 MHz, but its 1400 MHz traverse seems to be distorted by what may be a 90\degr\ ``jump'' just prior to the profile center. Its profile then seems to be a core/outer cone triple, and the quantitative beam geometry model is shown in Table~5. The core width is estimated around $2.65\degr$, which is much narrower than the central component of the profile ($7.92\degr$ at 327 MHz and $10.08\degr$ at 1400 MHz ); this indicates that the magnetic axis is canted at 42\degr $\pm 2.7\degr$. 

\begin{figure*}
\includegraphics[width=0.8\textwidth]{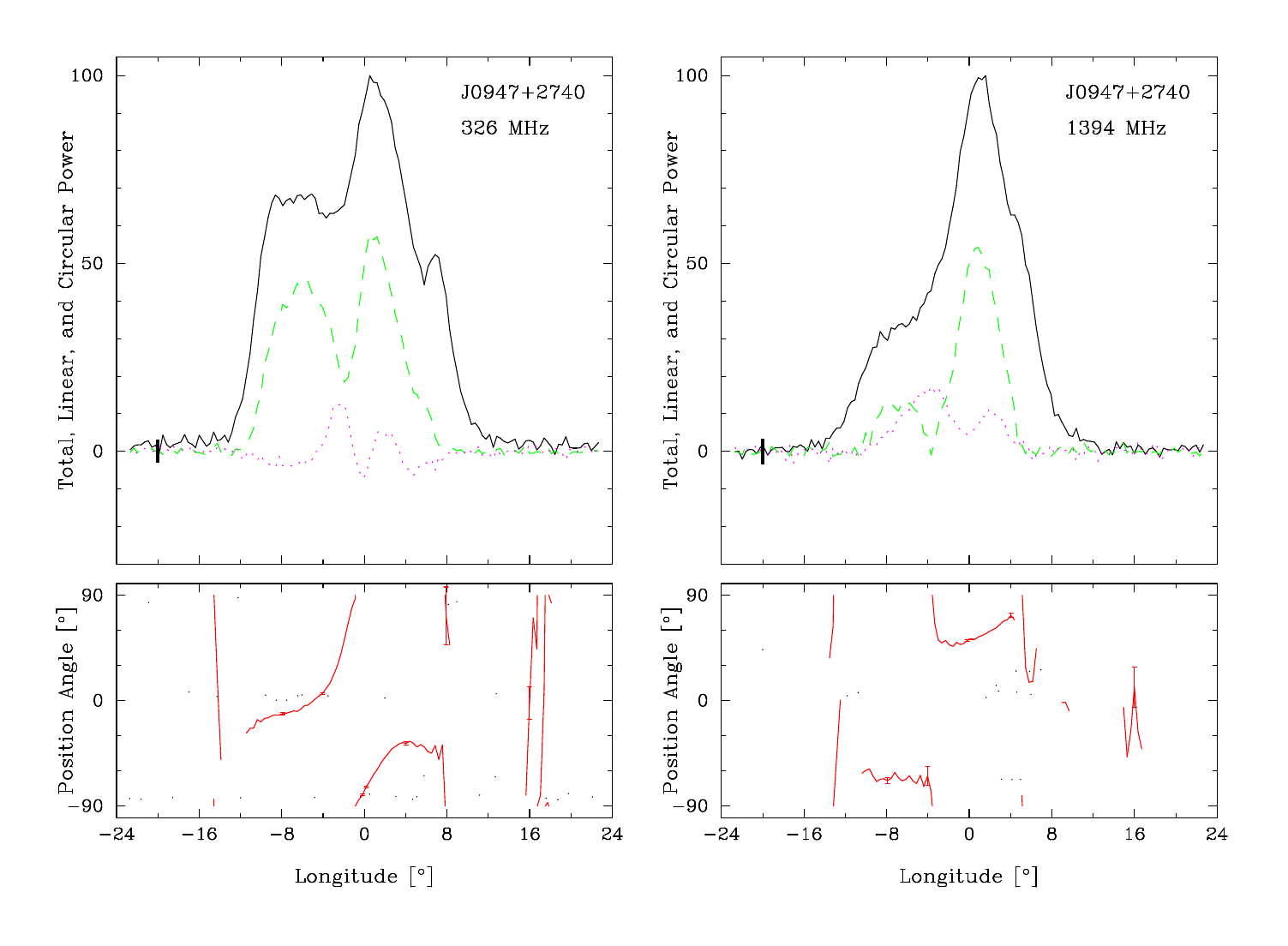}
\caption{J0947+2740 polarized profiles as in Fig.~\ref{figA1}.} 
\label{figA6}
\end{figure*}

\noindent{\bf J1246+2253} has a single component at 327 MHz, which develops into a 
resolved triple form at 1400 MHz in what may be the characteristic core-single manner.  
The fractional linear polarization at both frequencies is low, so little can be discerned reliably from the PPA tracks.  Also no clear features are seen in the fluctuation spectra as is often the case for core-single profiles.  

The single profile at 327 MHz becoming triple at 1400 MHz strongly suggests a core-single configuration.  Despite the low fractional linear polarization, the trailing positive traverse measured at 90\degr\ was used to compute the geometry in Table~5. The core width is calculated at $3.55\degr$. 

\begin{figure*}
\includegraphics[width=0.8\textwidth]{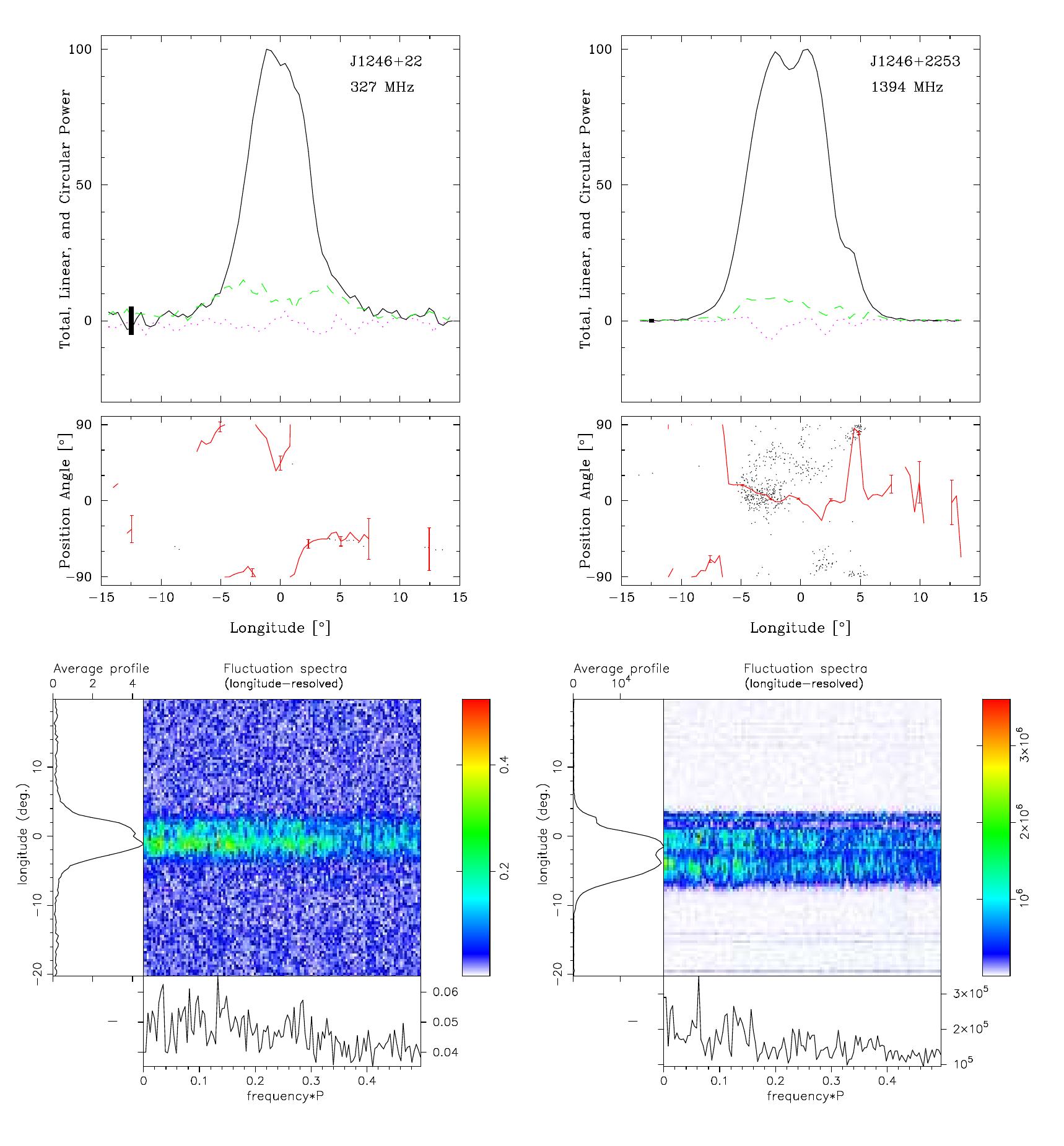}
\caption{J1246+2253 polarized profiles and fluctuation spectra as in Fig.~\ref{figA1}.} 
\label{figA7}
\end{figure*}

\begin{figure*}
\includegraphics[width=\textwidth]{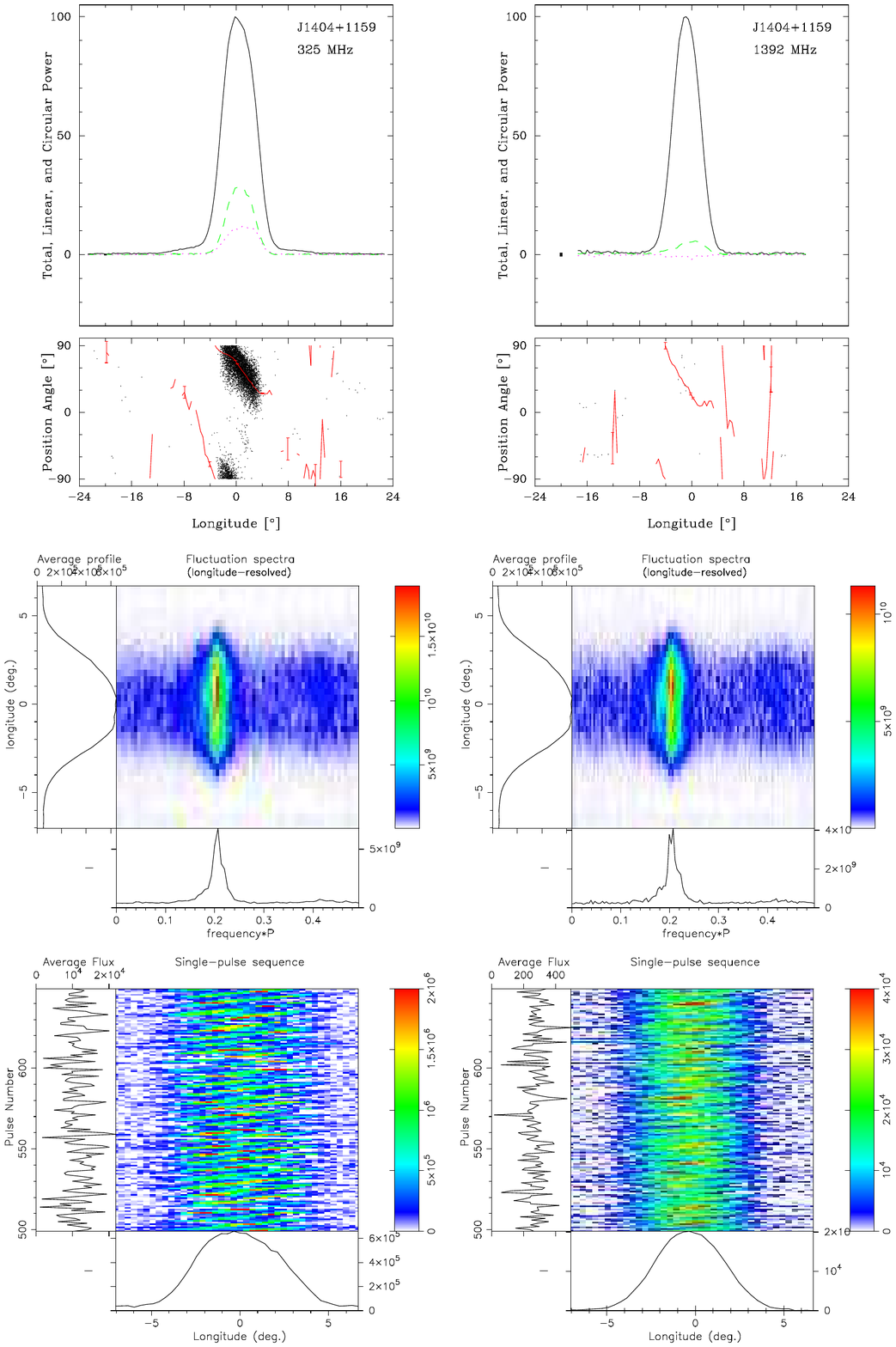}
\caption{J1404+1159 polarized profiles and fluctuation spectrum as in 
Fig.~\ref{figA1}. Additionally, a display showing the pulsar's accurately drifting 
subpulses is given along the bottom.}
\label{figA8}
\end{figure*}

\noindent{\bf J1404+1159} exhibits a narrow peak in its fluctuation spectra around 0.2 
cycles/period, suggesting a modulation period $P_3$ of some 5 rotation periods.  A display of its individual pulses bears this out, and a plot of its emission folded at $P_3$ shows that the modulation is highly regular.  The PPA rate at the profile center suggests an outside sightline traverse as is usual for conal single ``drifters''. See Figure~\ref{figA8}.  

The pulsar then appears to have a classic conal single profile.  Both PPA tracks show a 
negative-going traverse with a central slope of  --12\degr/\degr.  Here we also give  short pulse-sequences folded at the modulation period of some 0.2 rotation-periods/cycle.  

\noindent{\bf J1756+1822} appears to have two profile components, and its profile broadens perceptibly with wavelength, suggesting a conal double configuration. Both frequency profiles have similar forms, with the longer wavelength profile somewhat broader in the usual pattern of the conal double class.  The fractional linear profiles are low at both frequencies, but at 1400 MHz there seems to be a hint of a traverse through more than 90\degr\ across the profile. The fluctuation spectra 
are not shown as no features could be discerned. 

\begin{figure*}
\includegraphics[width=0.8\textwidth]{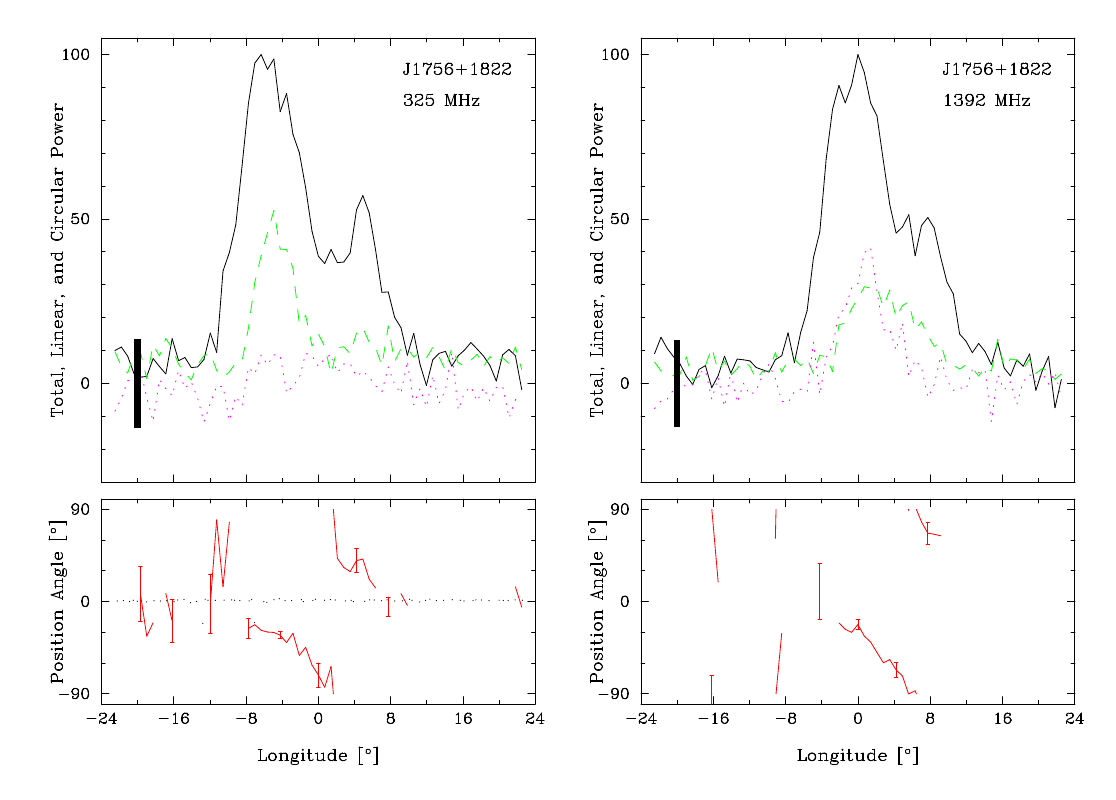}
\caption{J1756+1822 polarized profiles as in Fig.~\ref{figA1}.} 
\label{figA9}
\end{figure*}

\noindent{\bf J1935+1159}'s long nulls (see Fig.~\ref{fig2}) make this pulsar difficult to observe sensitively, and neither clear profile structure nor polarization signature is seen at either frequency. Similarly, fluctuation spectra showed nothing useful.  

This appears to be a four or five component pulsar even though only three components can be resolved, it appears to have the central PPA traverse and usual filled ``boxy'' form which could hide additional components. 

The emission heights, and the profile narrowing from 327 MHz to 1400 MHz, indicate that this is an outer cone which is filled with inner conal and/or core emission. For the geometric computations, we have estimated an unresolved central traverse of $10 \pm 2\degr/\degr$. 

\begin{figure*}
\includegraphics[width=0.8\textwidth]{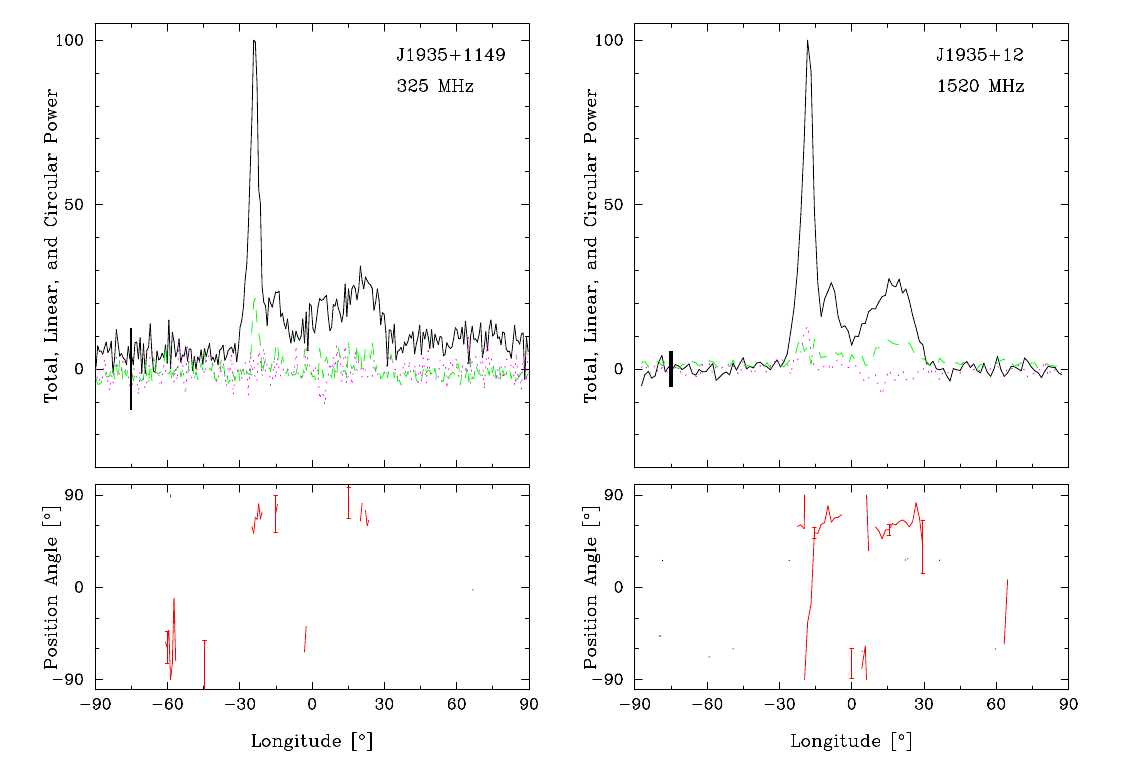}
\caption{J1935+1159 polarized profiles as in Fig.~\ref{figA1}.} 
\label{figA10}
\end{figure*}

\noindent{\bf J2050+1259} also exhibits frequent nulls as seen above in Fig.~\ref{fig2}.  However, its single profile at 1400 MHz broadens and bifurcates at 327 MHz in the usual conal double manner, and a steep PPA traverse is seen at the lower frequency. Here we do see strong low frequency features in the fluctuation spectra indicative of modulation on a scale of 50 rotation periods or longer. 

\begin{figure*}
\includegraphics[width=0.8\textwidth]{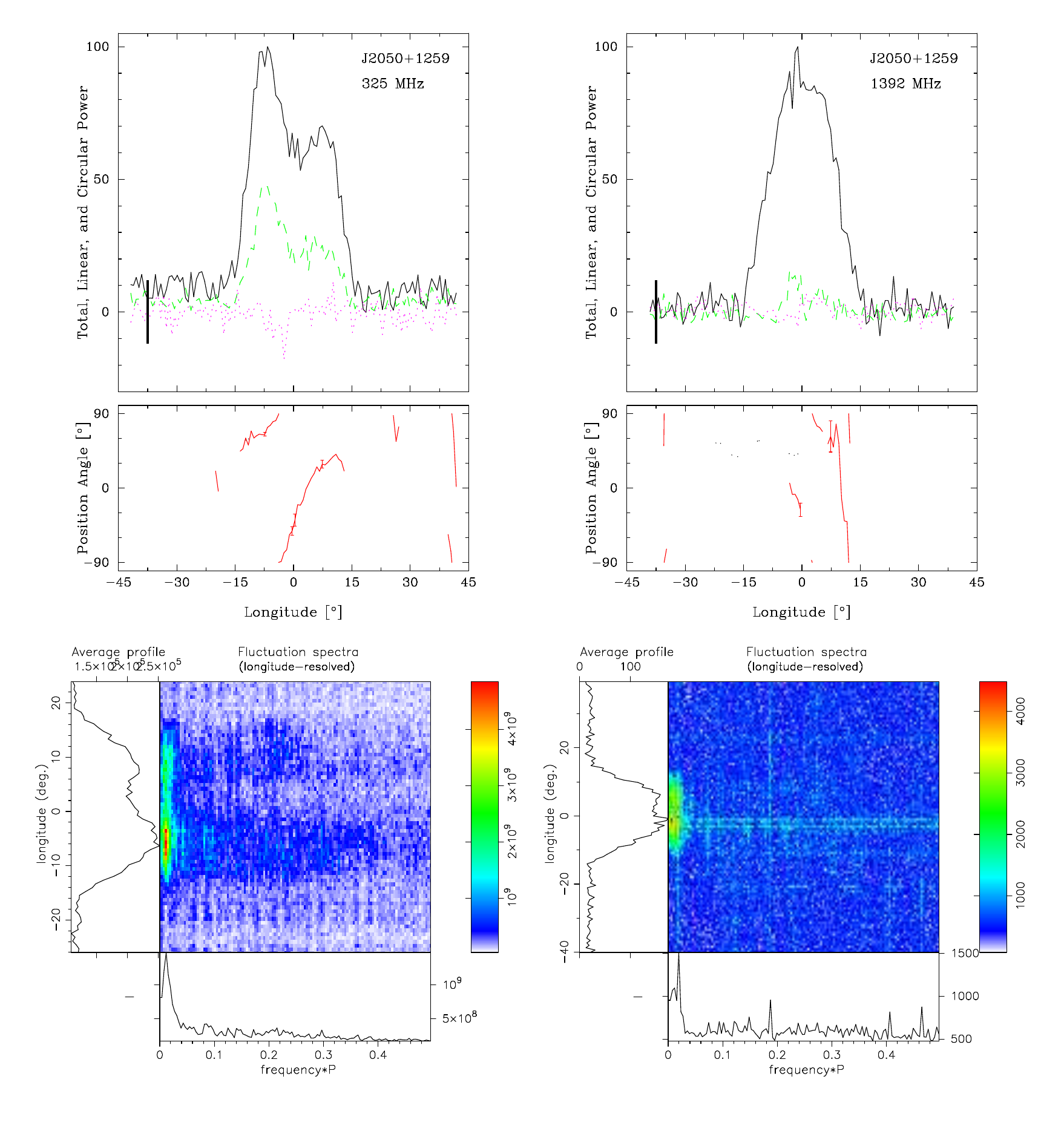}
\caption{J2050+1259 polarized profiles and fluctuation spectra as in  Fig.~\ref{figA1}.} 
\label{figA11}
\end{figure*}

\begin{figure*}
\includegraphics[width=0.8\textwidth]{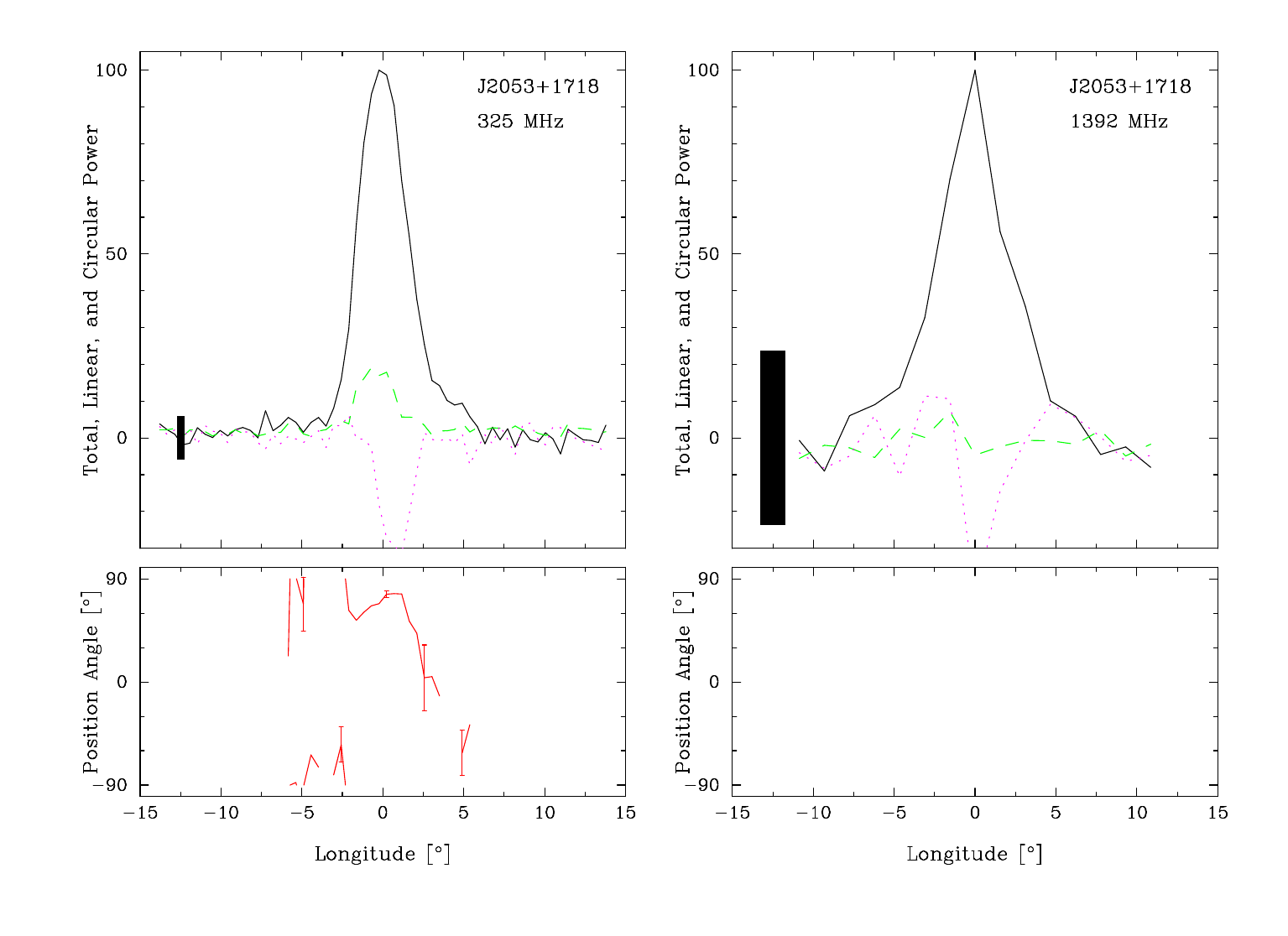}
\caption{J2053+1718 polarized profiles as in  Fig.~\ref{figA1}.} 
\label{figA12}
\end{figure*}

The profile has a single component at 1400 MHz and two closely spaced components at 327 MHz as seen in Figs.~\ref{fig2} and \ref{figA11}.  Its profile is broader at the lower frequency, and strong hints of a 90--180\degr\ PPA traverse are seen in both profiles, suggesting that this is a conal single profile with a small impact angle. Its beaming parameters are given in Table~5.  

\noindent{\bf J2053+1718}. Our observations do not provide much polarization information or fluctuation-spectral information apart from it having a single profile at both frequencies. In the higher frequency observation the time resolution was poor with only 232 samples across the rotation cycle. This pulsar has a wider profile at 1400 MHz than at 327 MHz; however, we hesitate to make conclusions about profile measurements due to the low quality of the observation. Despite this, its short 119-ms rotation period and single profile do suggest a core-single classification. 

\begin{table*} 
\begin{tabular}{|l| |l| |l| |l| |l| |l| |l| |c|} 
\hline 
 Name   & \multicolumn{2}{|c|}{MJD}                & \multicolumn{2}{|c|}{Number of Pulses}  & \multicolumn{2}{|c|}{Flux Density (mJy)} & RM \\ 
              & 327 MHz   & 1400 MHz & 327 MHz & 1400 MHz & 327 MHz & 1400 MHz & (rad-m$^2$) \\ 
\hline
J0435+2749 & 56327 & 54541  & 1830  & 3065  & 3.4(4)  & 0.24(3)  & +2 \\ 
J0517+2212 & 57123 & 54540  & 2696  & 2698  & 3.4(1.5)& 0.46(2)  & $-$16 \\ 
J0627+0706 & 57123 & 57113  & 518   & 2520  & 17(4)   & 1.0(1)   & +212 \\ 
J0927+2345 & 57525 & 57347  & 4724  & 2502  & 1.2(3)  & 0.06(1)  & $-$8 \\ 
J0943+2253 & 57379 & 57377  & 7775  & 13680 & 3.1(3)  & 0.39(7)  & +8? \\  
J0947+2740 & 57379 & 57524  & 3172  & 4699  & 2.3(3)  & 0.13(1)  & +32 \\ 
J1246+2253 & 56176 & 57307  & 1266  & 1885  & 2.4(7)  & 0.39(3)  & +4? \\ 
J1404+1159 & 55905 & 57307  & 509   & 1591  & 4.3(9)  & 0.027(5) & +4 \\
J1756+1822 & 57567 & 57567  & 2156  & 3571  &$<$0.002 & 0.014(5) & +70 \\ 
J1935+1159 & 57288 & 57533  & 1004  & 1090  & 1.5(9)  & 0.17(3)  & $-$83 \\ 
J2050+1259 & 57524 &57525   & 2046  & 1966  & 2.6(6)  & 0.05(1)  & $-$80 \\  
J2053+1718 & 57524 &57533   & 5023  & 21624 & 4(2)    & 0.003(2) & $-$5 \\ 
\hline 
\end{tabular} 
\caption{Polarimetric observations.}
\label{tab4}
\medskip 
\end{table*} 

\begin{table*}
\begin{tabular}{l c c c c c c c c c l}
\hline
Pulsar &Class& $w_{327}$ & $w_{1400}$ &  $\alpha$ &$\beta$ & $w$  &$\rho$ & $h$ & $R$ & Notes  \\
            &&  (\degr) & (\degr) & (\degr)& (\degr) &(\degr) &(\degr) &(km) & (\degr/\degr) & \\
\hline
J0435+2749 &T & 63.0 & 59.0 & 18 & +3.4 & 59.0 & 10.3 & 230 & 5 & \\
J0517+2212 &D& 55.0 & 45.5 & 31 & 0 &  45.2 &11.7 & 205 & 6 & Both frequencies have a flat PPA traverse, \\
& & & & & & & & & & suggesting central sightline, $\beta \, \sim \,$0\degr\ \\
J0627+0706m &S${_t}$& 4.5 & 4.5 & 90 & $-$7.2 & 4.5 & 7.5 & 131 & 8 & \\
J0627+0706i   & D  & 6.0 & 6.0 & 90 & $-$6.4 & 6 & 6.9 & 129 & 9 & $178\degr$ apart from main pulse.\\
J0927+2345m &D/T?& 5.0 & 6.4 & 69 & +3.4 & 8 & 5.1 & 130 & 16 & \\
J0943+2253 &D/T?& 6.5 & 7.8 & 42 & $-$5.5 & 7 & 5.9 & 123 & 7 & $R$ was measured at 327 MHz. \\
J0947+2740 &T/M& 19.4 & 17.0 & 42 & $-$1.2 & 18 & 6.0 & 205 & 32 & Core widths: $\sim$8\degr, 10\degr\ at 327, 1400 MHz\\
J1246+2253 &S${_t}$& 5.7 & 7.4 & 81 & $-$4.7 & 6.5 & 5.7 & 103 & 12 & \\
J1404+1159 &S${_d}$& 5.8 & 5.0 & 25 & +2.2 & 5.3 & 0.88 & 114 & 11 & \\
J1756+1822 &D& 16.0 & 15.3 & 50 & +2.5 & 16 & 6.7 & 224 & 18 & Core widths: $8.2\degr$, $7.1\degr$ at 327, 1400 MHz. \\
& & & & & & & & & & $R$ was measured at 1400 MHz.\\
J1935+1159 &D & 51.8 & 50.5 & 10.13 & $-$1.0 & 50.4 & 4.3 & 239 & 10 & \\
J2050+1259 &S${_d}$& 25.0 & 18.0 & 27 & $-$2.6 & 21 & 5.2 & 223 & 10 & $R$ measured at 327 MHz\\
J2053+1718 &S${_t}$?& 2.5 & 5.0 & 63 & $-$17.2 & 2.5 & 17.2 & 235 & 16 & \\
\hline
\end{tabular}
\caption{Conal Geometry Models. The profile classes are defined in Rankin (1993a). $w_{327}$
and $w_{1400}$ represent the half-power pulse widths (in degrees) at 327 and 1400 MHz respectively.
$\alpha$ values are estimated from core component widths per ET VIa, eq.(1) where possible and $\beta$ from eq.(3). The outer half-power widths $w$ are interpolated to 1 GHz from profile measurements, and then $\rho$ and $h$ computed using eqs.(4) and (6). $R$ is the polarization position angle slope. (a) No data available for PSR J0943+2253 at 1400 MHz.}
\label{tab5}
\end{table*}

\section{On the origin of PSR J2053+1718}
\label{2053+1718}

In double neutron star systems, the first-born neutron star was recycled by accretion of mass
from the progenitor of the second-formed neutron star. This accretion spun up the
first formed NS to spin periods between 22 and 186 ms (for the known DNSs, likely
faster right after they formed) and, by mechanisms that are not clearly understood,
it induced a decrease in its magnetic field to values between $10^9\, <\, B_0 \, <
\, 10^{10}$ G. Such pulsars spin down very slowly and therefore will stay in the
active part of the $P$ - $\dot{P}$ diagram for much longer than non-recycled
pulsars; this is the reason why we mostly see the recycled pulsars in these systems
(the second-formed non-recycled pulsars are observed in two DNSs only). For a review
see \cite{2017ApJ...846..170T}.

Some isolated pulsars like PSR~J2053+1718
 are in the same area of the $P$ - $\dot{P}$
diagram as the recycled pulsars in DNSs, but have no companion to explain the
recycling. The conventional explanation for their formation is that, like the
recycled pulsars in DNSs, they were spun up by a massive stellar companion; the
difference is that when the latter star goes supernova and forms a neutron star, the system unbinds,
owing to the kick and mass loss associated with the supernova (e.g.,
\citealt{2010MNRAS.407.1245B}). For this reason they have been labelled 
``Disrupted Recycled Pulsars'', or DRPs.

We should keep in mind the possibility of alternative origins for these pulsars:
some NSs observed in the center of supernova remnants, despite being obviously
young, have small B-fields and large characteristic ages similar to those of DRPs
--- these objects are known as Central Compact Objects, or CCOs
(see e.g., \citealt{2010ApJ...709..436H}). Therefore,
one could expect that some DRPs formed as CCOs. However, \cite{2013ApJ...773..141G}
find after extensive study of DRPs in X-rays that none appears to have thermal
X-ray emission, implying that there is likely no relation between CCOs and DRPs.
This results in a mystery: a substantial fraction of neutron stars appear to form as
CCOs, which one might reasonably expect to form DRP-like radio pulsars, some of them
with strong thermal X-ray emission, but these large numbers of DRPs (and ``hot''
DRPs) are not observed. Either, for some unknown reason, they never develop radio
emission, or their B-field increases significantly after birth, making them look
more like normal pulsars. In any case, the results of \cite{2013ApJ...773..141G}
seem to exclude the possibility that DRPs such as PSR~J2053+1718
formed from central Compact Objects.

Any model that explains quantitatively the observed distribution of
orbital eccentricities
and spatial velocities of DNSs should also be able to explain the relative fraction
of DRPs to DNSs and, furthermore, the velocity distribution of the two classes
of objects.  As already discussed by \cite{2010MNRAS.407.1245B}, the relative number of
DNSs and DRPs implies that the second SN kick must be, on average,
significantly smaller than that observed for single pulsars. The evidence for this
among DNSs has been growing in recent years and is now very strong (see comprehensive
discussion in
\citealt{2017ApJ...846..170T}). It is therefore clear that more measurements of proper motions,
spin periods, ages and B-fields of DRPs such as those presented here give us important
clues for understanding the formation of DNS systems.

\section{Conclusions} 
\label{disc}

In the foregoing sections we have characterized a 
group of pulsars that had not been the target of any previous detailed studies.
We determined timing solutions for them; most are normal, isolated pulsars.
One of them, PSR~J0627+0706, is relatively young ($\tau_c \, = \, 250$ kyr)
and is located near the Monoceros SNR, however it is unlikely that the
pulsar and that SNR originated in the same supernova event: We find that a nearby
``radio quiet'' pulsar recently discovered by the Fermi satellite, PSR~J0633+0632,
is located closer to the centre of the cluster and has a characteristic
age that agrees better with the estimated age of the SNR.

We confirmed a candidate from a previous survey, PSR J2053+1718;
subsequent timing shows that, despite being solitary, this object was recycled;
it appears to be a member of a growing class of objects that appear to result from
the disruption of double neutron stars at formation. We highlight that measurements
of the characteristics of these objects (spin period, age, B-field, velocity)
are important for understanding the formation of double neutron star systems.

As part of our characterization, we have also observed these pulsars polarimetrically with the Arecibo telescope at both 1400 MHz and 327 MHz in an effort to explore their pulse-sequence 
properties and quantitative geometry.  
 Three of them (PSRs~J0943+2253, J1935+1159 and J2050+1259) 
have strong nulls and sporadic radio emission, several others exhibit interpulses (PSRs 
J0627+0706 and J0927+2345) and one shows regular drifting subpulses (J1404+1159).
All these measurements will contribute to future, more global assessments of the
emission properties of radio pulsars and studies of the NS population in the Galaxy.

\section*{Acknowledgments}  Much of the work was made possible by support from the US National Science Foundation grant 09-68296 as well as NASA Space Grants.  One of us (JMR) also thanks the Anton Pannekoek Astronomical Institute of the University of Amsterdam for their support.  Arecibo Observatory is operated by SRI International under a cooperative agreement with the US National Science Foundation, and in alliance with SRI, the Ana G. M\'endez-Universidad Metropolitana, and the Universities Space Research Association.  This work made use of the NASA 
ADS astronomical data system. One of us (KS) thanks the US National Science Foundation for their support under the Physics Frontiers Center award number 1430284.





\begin{thebibliography}{}
\makeatletter
\relax
\def\mn@urlcharsother{\let\do\@makeother \do\$\do\&\do\#\do\^\do\_\do\%\do\~}
\def\mn@doi{\begingroup\mn@urlcharsother \@ifnextchar [ {\mn@doi@}
  {\mn@doi@[]}}
\def\mn@doi@[#1]#2{\def\@tempa{#1}\ifx\@tempa\@empty \href
  {http://dx.doi.org/#2} {doi:#2}\else \href {http://dx.doi.org/#2} {#1}\fi
  \endgroup}
\def\mn@eprint#1#2{\mn@eprint@#1:#2::\@nil}
\def\mn@eprint@arXiv#1{\href {http://arxiv.org/abs/#1} {{\tt arXiv:#1}}}
\def\mn@eprint@dblp#1{\href {http://dblp.uni-trier.de/rec/bibtex/#1.xml}
  {dblp:#1}}
\def\mn@eprint@#1:#2:#3:#4\@nil{\def\@tempa {#1}\def\@tempb {#2}\def\@tempc
  {#3}\ifx \@tempc \@empty \let \@tempc \@tempb \let \@tempb \@tempa \fi \ifx
  \@tempb \@empty \def\@tempb {arXiv}\fi \@ifundefined
  {mn@eprint@\@tempb}{\@tempb:\@tempc}{\expandafter \expandafter \csname
  mn@eprint@\@tempb\endcsname \expandafter{\@tempc}}}

\bibitem[\protect\citeauthoryear{{Abdo} et~al.,}{{Abdo}
  et~al.}{2009}]{2009Sci...325..840A}
{Abdo} A.~A.,  et~al., 2009, \mn@doi [Science] {10.1126/science.1175558}, \href
  {http://adsabs.harvard.edu/abs/2009Sci...325..840A} {325, 840}

\bibitem[\protect\citeauthoryear{{Belczynski}, {Lorimer}, {Ridley}  \&
  {Curran}}{{Belczynski} et~al.}{2010}]{2010MNRAS.407.1245B}
{Belczynski} K.,  {Lorimer} D.~R.,  {Ridley} J.~P.,   {Curran} S.~J.,  2010,
  \mn@doi [\mnras] {10.1111/j.1365-2966.2010.16970.x}, \href
  {http://adsabs.harvard.edu/abs/2010MNRAS.407.1245B} {407, 1245}

\bibitem[\protect\citeauthoryear{{Burgay} et~al.,}{{Burgay}
  et~al.}{2013}]{2013MNRAS.429..579B}
{Burgay} M.,  et~al., 2013, \mn@doi [\mnras] {10.1093/mnras/sts359}, \href
  {http://adsabs.harvard.edu/abs/2013MNRAS.429..579B} {429, 579}

\bibitem[\protect\citeauthoryear{{Chandler}}{{Chandler}}{2003}]{2003PhDT.........2C}
{Chandler} A.~M.,  2003, PhD thesis, California Institute of Technology

\bibitem[\protect\citeauthoryear{{Cordes} \& {Lazio}}{{Cordes} \&
  {Lazio}}{2002}]{2002astro.ph..7156C}
{Cordes} J.~M.,  {Lazio} T.~J.~W.,  2002, arXiv:astro-ph/0207156, \href{http://adsabs.harvard.edu/abs/2002astro.ph..7156C} { }

\bibitem[\protect\citeauthoryear{{Damour} \& {Taylor}}{{Damour} \&
  {Taylor}}{1991}]{1991ApJ...366..501D}
{Damour} T.,  {Taylor} J.~H.,  1991, \mn@doi [\apj] {10.1086/169585}, \href
  {http://adsabs.harvard.edu/abs/1991ApJ...366..501D} {366, 501}

\bibitem[\protect\citeauthoryear{{Demorest} et~al.,}{{Demorest}
  et~al.}{2013}]{2013ApJ...762...94D}
{Demorest} P.~B.,  et~al., 2013, \mn@doi [\apj] {10.1088/0004-637X/762/2/94},
  \href {http://adsabs.harvard.edu/abs/2013ApJ...762...94D} {762, 94}

\bibitem[\protect\citeauthoryear{{Deneva}, {Stovall}, {McLaughlin}, {Bates},
  {Freire}, {Martinez}, {Jenet}  \& {Bagchi}}{{Deneva}
  et~al.}{2013}]{2013ApJ...775...51D}
{Deneva} J.~S.,  {Stovall} K.,  {McLaughlin} M.~A.,  {Bates} S.~D.,  {Freire}
  P.~C.~C.,  {Martinez} J.~G.,  {Jenet} F.,   {Bagchi} M.,  2013, \mn@doi
  [\apj] {10.1088/0004-637X/775/1/51}, \href
  {http://adsabs.harvard.edu/abs/2013ApJ...775...51D} {775, 51}

\bibitem[\protect\citeauthoryear{{Deshpande} \& {Rankin}}{{Deshpande} \&
  {Rankin}}{2001}]{2001MNRAS.322..438D}
{Deshpande} A.~A.,  {Rankin} J.~M.,  2001, \mn@doi [\mnras]
  {10.1046/j.1365-8711.2001.04079.x}, \href
  {http://adsabs.harvard.edu/abs/2001MNRAS.322..438D} {322, 438}

\bibitem[\protect\citeauthoryear{{Dowd}, {Sisk}  \& {Hagen}}{{Dowd}
  et~al.}{2000}]{2000ASPC..202..275D}
{Dowd} A.,  {Sisk} W.,   {Hagen} J.,  2000, in {Kramer} M.,  {Wex} N.,
  {Wielebinski} R.,  eds,  Astronomical Society of the Pacific Conference
  Series Vol. 202, IAU Colloq. 177: Pulsar Astronomy - 2000 and Beyond. pp
  275--276

\bibitem[\protect\citeauthoryear{{Gonzalez} et~al.,}{{Gonzalez}
  et~al.}{2011}]{2011ApJ...743..102G}
{Gonzalez} M.~E.,  et~al., 2011, \mn@doi [\apj] {10.1088/0004-637X/743/2/102},
  \href {http://adsabs.harvard.edu/abs/2011ApJ...743..102G} {743, 102}

\bibitem[\protect\citeauthoryear{{Gotthelf}, {Halpern}, {Allen}  \&
  {Knispel}}{{Gotthelf} et~al.}{2013}]{2013ApJ...773..141G}
{Gotthelf} E.~V.,  {Halpern} J.~P.,  {Allen} B.,   {Knispel} B.,  2013, \mn@doi
  [\apj] {10.1088/0004-637X/773/2/141}, \href
  {http://adsabs.harvard.edu/abs/2013ApJ...773..141G} {773, 141}

\bibitem[\protect\citeauthoryear{{Halpern} \& {Gotthelf}}{{Halpern} \&
  {Gotthelf}}{2010}]{2010ApJ...709..436H}
{Halpern} J.~P.,  {Gotthelf} E.~V.,  2010, \mn@doi [\apj]
  {10.1088/0004-637X/709/1/436}, \href
  {http://adsabs.harvard.edu/abs/2010ApJ...709..436H} {709, 436}

\bibitem[\protect\citeauthoryear{{Hewish}, {Bell}, {Pilkington}, {Scott}  \&
  {Collins}}{{Hewish} et~al.}{1968}]{1968Natur.217..709H}
{Hewish} A.,  {Bell} S.~J.,  {Pilkington} J.~D.~H.,  {Scott} P.~F.,   {Collins}
  R.~A.,  1968, \mn@doi [\nat] {10.1038/217709a0}, \href
  {http://adsabs.harvard.edu/abs/1968Natur.217..709H} {217, 709}

\bibitem[\protect\citeauthoryear{{Leahy}, {Naranan}  \& {Singh}}{{Leahy}
  et~al.}{1986}]{1986MNRAS.220..501L}
{Leahy} D.~A.,  {Naranan} S.,   {Singh} K.~P.,  1986, \mn@doi [\mnras]
  {10.1093/mnras/220.3.501}, \href
  {http://adsabs.harvard.edu/abs/1986MNRAS.220..501L} {220, 501}

\bibitem[\protect\citeauthoryear{{Lorimer} \& {Kramer}}{{Lorimer} \&
  {Kramer}}{2004}]{2004hpa..book.....L}
{Lorimer} D.~R.,  {Kramer} M.,  2004, {Handbook of Pulsar Astronomy}

\bibitem[\protect\citeauthoryear{{Lorimer}, {Camilo}  \&
  {McLaughlin}}{{Lorimer} et~al.}{2013}]{2013MNRAS.434..347L}
{Lorimer} D.~R.,  {Camilo} F.,   {McLaughlin} M.~A.,  2013, \mn@doi [\mnras]
  {10.1093/mnras/stt1023}, \href
  {http://adsabs.harvard.edu/abs/2013MNRAS.434..347L} {434, 347}

\bibitem[\protect\citeauthoryear{{Manchester}, {Hobbs}, {Teoh}  \&
  {Hobbs}}{{Manchester} et~al.}{2005}]{2005AJ....129.1993M}
{Manchester} R.~N.,  {Hobbs} G.~B.,  {Teoh} A.,   {Hobbs} M.,  2005, \mn@doi
  [\aj] {10.1086/428488}, \href
  {http://adsabs.harvard.edu/abs/2005AJ....129.1993M} {129, 1993}

\bibitem[\protect\citeauthoryear{{Mitra}, {Rankin}  \& {Arjunwadkar}}{{Mitra}
  et~al.}{2016}]{2016MNRAS.460.3063M}
{Mitra} D.,  {Rankin} J.,   {Arjunwadkar} M.,  2016, \mn@doi [\mnras]
  {10.1093/mnras/stw1186}, \href
  {http://adsabs.harvard.edu/abs/2016MNRAS.460.3063M} {460, 3063}

\bibitem[\protect\citeauthoryear{{Navarro}, {Anderson}  \& {Freire}}{{Navarro}
  et~al.}{2003}]{2003ApJ...594..943N}
{Navarro} J.,  {Anderson} S.~B.,   {Freire} P.~C.,  2003, \mn@doi [\apj]
  {10.1086/377153}, \href {http://adsabs.harvard.edu/abs/2003ApJ...594..943N}
  {594, 943}

\bibitem[\protect\citeauthoryear{{Odegard}}{{Odegard}}{1986}]{1986ApJ...301..813O}
{Odegard} N.,  1986, \mn@doi [\apj] {10.1086/163945}, \href
  {http://adsabs.harvard.edu/abs/1986ApJ...301..813O} {301, 813}

\bibitem[\protect\citeauthoryear{{Rankin}}{{Rankin}}{1993a}]{1993ApJS...85..145R}
{Rankin} J.~M.,  1993a, \mn@doi [\apjs] {10.1086/191758}, \href
  {http://adsabs.harvard.edu/abs/1993ApJS...85..145R} {85, 145} (ET VIa)

\bibitem[\protect\citeauthoryear{{Rankin}}{{Rankin}}{1993b}]{1993ApJ...405..285R}
{Rankin} J.~M.,  1993b, \mn@doi [\apj] {10.1086/172361}, \href
  {http://adsabs.harvard.edu/abs/1993ApJ...405..285R} {405, 285} (ET VIb)

\bibitem[\protect\citeauthoryear{{Rankin}, {Wright}  \& {Brown}}{{Rankin}
  et~al.}{2013}]{2013MNRAS.433..445R}
{Rankin} J.~M.,  {Wright} G.~A.~E.,   {Brown} A.~M.,  2013, \mn@doi [\mnras]
  {10.1093/mnras/stt739}, \href
  {http://adsabs.harvard.edu/abs/2013MNRAS.433..445R} {433, 445}

\bibitem[\protect\citeauthoryear{{Ransom}, {Eikenberry}  \&
  {Middleditch}}{{Ransom} et~al.}{2002}]{2002AJ....124.1788R}
{Ransom} S.~M.,  {Eikenberry} S.~S.,   {Middleditch} J.,  2002, \mn@doi [\aj]
  {10.1086/342285}, \href {http://adsabs.harvard.edu/abs/2002AJ....124.1788R}
  {124, 1788}

\bibitem[\protect\citeauthoryear{{Ray}, {Thorsett}, {Jenet}, {van Kerkwijk},
  {Kulkarni}, {Prince}, {Sandhu}  \& {Nice}}{{Ray}
  et~al.}{1996}]{1996ApJ...470.1103R}
{Ray} P.~S.,  {Thorsett} S.~E.,  {Jenet} F.~A.,  {van Kerkwijk} M.~H.,
  {Kulkarni} S.~R.,  {Prince} T.~A.,  {Sandhu} J.~S.,   {Nice} D.~J.,  1996,
  \mn@doi [\apj] {10.1086/177934}, \href
  {http://adsabs.harvard.edu/abs/1996ApJ...470.1103R} {470, 1103}

\bibitem[\protect\citeauthoryear{{Ray} et~al.,}{{Ray}
  et~al.}{2011}]{2011ApJS..194...17R}
{Ray} P.~S.,  et~al., 2011, \mn@doi [\apjs] {10.1088/0067-0049/194/2/17}, \href
  {http://adsabs.harvard.edu/abs/2011ApJS..194...17R} {194, 17}

\bibitem[\protect\citeauthoryear{{Reid} et~al.,}{{Reid}
  et~al.}{2014}]{2014ApJ...783..130R}
{Reid} M.~J.,  et~al., 2014, \mn@doi [\apj] {10.1088/0004-637X/783/2/130},
  \href {http://adsabs.harvard.edu/abs/2014ApJ...783..130R} {783, 130}

\bibitem[\protect\citeauthoryear{{Shklovskii}}{{Shklovskii}}{1970}]{1970SvA....13..562S}
{Shklovskii} I.~S.,  1970, \sovast, \href
  {http://adsabs.harvard.edu/abs/1970SvA....13..562S} {13, 562}

\bibitem[{{van Straten} {et~al.}(2012){van Straten}, {Demorest}, \&
  {Oslowski}}]{2012AR&T....9..237V}
{van Straten}, W., {Demorest}, P., \& {Oslowski}, S. 2012,
  Astronomical Research and Technology, {9, 237}

\bibitem[\protect\citeauthoryear{{Tauris} et~al.,}{{Tauris}
  et~al.}{2017}]{2017ApJ...846..170T}
{Tauris}, T.~M., Kramer, M., Freire, P.~C.~C., et al. 2017, \mn@doi [\apj] {10.3847/1538-4357/aa7e89}, \href
{http://adsabs.harvard.edu/abs/2017ApJ...846..170T} {846, 170}

\bibitem[\protect\citeauthoryear{{Taylor}}{{Taylor}}{1992}]{1992RSPTA.341..117T}
{Taylor} J.~H.,  1992, \mn@doi [Philosophical Transactions of the Royal Society
  of London Series A] {10.1098/rsta.1992.0088}, \href
  {http://adsabs.harvard.edu/abs/1992RSPTA.341..117T} {341, 117}

\bibitem[\protect\citeauthoryear{{Thorsett}, {Deich}, {Kulkarni}, {Navarro}  \&
  {Vasisht}}{{Thorsett} et~al.}{1993}]{1993ApJ...416..182T}
{Thorsett} S.~E.,  {Deich} W.~T.~S.,  {Kulkarni} S.~R.,  {Navarro} J.,
  {Vasisht} G.,  1993, \mn@doi [\apj] {10.1086/173224}, \href
  {http://adsabs.harvard.edu/abs/1993ApJ...416..182T} {416, 182}

\bibitem[\protect\citeauthoryear{{Yao}, {Manchester}  \& {Wang}}{{Yao}
  et~al.}{2017}]{2017ApJ...835...29Y}
{Yao} J.~M.,  {Manchester} R.~N.,   {Wang} N.,  2017, \mn@doi [\apj]
  {10.3847/1538-4357/835/1/29}, \href
  {http://adsabs.harvard.edu/abs/2017ApJ...835...29Y} {835, 29}

\bibitem[\protect\citeauthoryear{{Young} \& {Rankin}}{{Young} \&
  {Rankin}}{2012}]{2012MNRAS.424.2477Y}
{Young} S.~A.~E.,  {Rankin} J.~M.,  2012, \mn@doi [\mnras]
  {10.1111/j.1365-2966.2012.21077.x}, \href
  {http://adsabs.harvard.edu/abs/2012MNRAS.424.2477Y} {424, 2477}

\makeatother
\end{thebibliography}

\newpage



\bsp	
\label{lastpage}
\end{document}